\documentclass[12pt]{article}
\usepackage[margin=1in]{geometry}

\usepackage{amsmath,amsthm,amsfonts,amssymb,mathrsfs}
\usepackage{enumitem}
\usepackage[colorlinks=true, pdfstartview=FitV, linkcolor=blue,citecolor=blue, urlcolor=blue]{hyperref}

\usepackage{times}

\usepackage{esint}

\usepackage{mathabx}
\usepackage{xfrac}
\usepackage[authoryear,round]{natbib}

\allowdisplaybreaks

\newtheoremstyle{thmboldstyle}
   {}{}{\itshape}{}{\bfseries}{.}{.5em}{{\thmname{#1 }}{\thmnumber{#2}}{\thmnote{ (#3)}}}
\theoremstyle{thmboldstyle}

\newtheoremstyle{remboldstyle}
   {}{}{}{}{\bfseries}{.}{.5em}{{\thmname{#1 }}{\thmnumber{#2}}{\thmnote{ (#3)}}}
\theoremstyle{remboldstyle}



\theoremstyle{thmboldstyle}

\theoremstyle{remboldstyle}

\numberwithin{equation}{section}

\usepackage{fancyhdr}
\def\p{\partial}
\def\jump#1{{[\hspace{-2pt}[#1]\hspace{-2pt}]}}

\def\bnabla{\boldsymbol\nabla}
\def\bcdot{\boldsymbol\cdot}

\title{{\bf 3D Interface Models for Rayleigh-Taylor Problems \\ \,
}}

\author{
{ \small {\bf Gavin Pandya}}\thanks{\footnotesize Department of Mathematics, UC Davis, Davis, CA 95616 \href{gpandya@ucdavis.edu}
 \href{gpandya@ucdavis.edu}{gpandya@ucdavis.edu}.}
\and  
{\small {\bf Steve  Shkoller}}\thanks{\footnotesize Department of Mathematics, UC Davis, Davis, CA 95616, \href{shkoller@math.ucdavis.edu}{shkoller@math.ucdavis.edu}.}
}

\date{}

\begin{document}
\maketitle

\begin{abstract}
We derive interface models for 3D Rayleigh-Taylor instability (RTI),  making use of a novel asymptotic expansion in the non-locality of the fluid flow.   These interface models are derived for the purpose of studying universal features associated to RTI such as the Froude number in single-mode RTI, the predicted quadratic growth of the interface amplitude under multi-mode random perturbations, the optimal (viscous) mixing rates induced by the RTI and the self-similarity of  horizontally averaged density profiles, and the remarkable stabilization of the mixing layer growth rate which arises for the three-fluid two-interface heavy-light-heavy configuration, in which the addition of a third fluid bulk slows the growth of the mixing layer to  a linear rate.   Our interface models can capture the formation of small-scale structures induced by severe interface roll-up, reproduce  experimental data in a number of different regimes, and study the effects of multiple interface interactions even as the interface separation  distance becomes exceedingly small.  Compared to traditional numerical schemes used to study such phenomena, our models provide a  computational speed-up of at least two orders of magnitude.
\end{abstract}

\section{Introduction}

Rayleigh-Taylor instability (RTI) occurs when the interface between two fluids of different density is subjected to a normal pressure gradient. 
When a fluid undergoes RTI, vorticity is deposited on the  material interface separating the two fluids.
This interface vorticity then initiates the Kelvin-Helmholtz instability (KHI), which causes the interface to roll-up into extremely complex shapes. 
RTI is fundamental to a wide variety of complex physical processes, and is often coupled to the effects of electromagnetism, gravity, reaction chemistry, or combustion, as well as the effects of shock waves and Richtmeyer-Meshkov instability. For a comprehensive review of the state of the art in the theory, computation, and application of RTI, see the two-part review of \citet{Zh2017-2,Zh2017-1}.

Our objective is the analysis and quantification of certain universal aspects of RTI by a novel derivation and implementation of  reduced-order interface models.   
Specifically, we wish to make use of the dimensional reduction afforded by irrotational,  incompressible, and inviscid fluids, coupled to an asymptotic expansion in the non-locality of the flow to derive accurate models of complex interface motion, which allow for roll-up, small-scale structure formation, multiple interface interaction and stratification, and viscous mixing via ensemble averaging of large sets of randomly initialized inviscid RTI runs.  
There are certain universal features associated with the RT instability that have been discovered experimentally, derived analytically by turbulence closure models, or established via Direct Numerical Simulation (DNS) which are fundamental to our understanding of turbulence. 
We are particularly interested in the universality  of  (a) the Froude number for bubble growth under a single-mode RT instability which, in turn, 
implies that the asymptotic bubble velocity depends only bubble size;  (b) the theoretically and experimentally predicted quadratic growth rate of
the interface amplitude under random multi-mode perturbations, (c) the optimal (viscous) mixing rates induced by the RT instability, and the
self-similar evolution of horizontally averaged density profiles; (d) the persistent formation of small-scale
features due to multiple fluid interface interactions; and finally (e) the remarkable self-similar stabilization of the mixing layer growth rate which arises
for the three-fluid two-interface heavy-light-heavy configuration, in which the addition of a third fluid bulk slows the growth of the mixing layer to a linear rate (as opposed to the standard quadratic growth rate for two fluid RTI).

Historically, early studies
of RTI focused on the development of single-mode perturbations of a flat interface, and studied questions concerning the nonlinear growth of these perturbations and their longtime behavior \citep{JaWi2007}. 
For instance, the asymptotic bubble/spike velocity of single-mode RTI appears to be a universal feature of bubbly flows, depending only on the size of the bubble, although there is some evidence of ``re-acceleration'' of bubbles at very late times \citep{CaDiFrRaYo2006}.
For 	many  engineering and physics applications, 
the  quantification of the growth of the mixing-layer  and  the rate of fluid mixing in RTI-driven turbulence is of fundamental importance \citep{Sh1984}. 
For example, the growth of the mixing layer in turbulent RTI has been an area of intense study for decades, starting with \citet{Re1984}. 
The growth of the interface amplitude  is the natural quantity to measure in a variety of experimental settings, exhibiting a universal quadratic growth rate at long times due to nonlinear mode interactions, as well as a self-similar scaling of the density profile \cite{BoLiMu2010}.
However, interesting questions remain about the dependence of the non-dimensional growth-rate parameter  upon the spectrum of the initial perturbation \citep{ClRi2004}, so that a correct prediction of this growth rate is essential to the study of RTI.
In practical terms, this predicted amplitude growth rate is a measure of the growth of the mixing-layer in the case that fluids have viscosity.  
It is of paramount interest to understand how fast multiple fluids {\it mix} under the action the RT interface instability. 
This  rate can be quantified by the  so-called ``mix norms'' which are defined in \citet{DoLiTh2011}). Therein,  it was shown that  for passive
scalar  transport by a specially constructed (shear-type) transport velocity, optimal mixing occurs in the sense of exponentially fast convergence to the scalar average.
While no theorems as yet exist for the case of mixing by the Navier-Stokes equations (or any other fluids equations for that matter), there is also very
little numerical evidence of optimal mixing under RTI.   The development of our reduced-order interface models is intended for the study of these 
fundamental behaviors.

Well-known reduced-order models which have been used to study the interface growth rate under RTI are the Haan and Goncharov models, which approximate single-mode and multi-mode initial perturbations, respectively \citep{Ha1989,Go2002}. 
All the aforementioned models assume that the interface remains a graph; as such, these models have limited
accuracy and predictive capabilities beyond the linear stage of the interface motion \citep[see][]{AnRo2013}.

Herein, we shall derive interface models for the  Rayleigh-Taylor instability by a suitable approximation of the multi-phase irrotational and incompressible Euler equations.  
Specifically, with the aim of preserving the geometric complexity of RTI, we do not use traditional  asymptotic assumptions of smallness  in either amplitude or slopes. 
Instead, we use an asymptotic expansion of the interaction potential between interfacial particle positions to derive  reduced-order approximations of the Euler equations, with a general parametrization  which allows for interface turnover and roll-up. 
This process results in three interface models for Rayleigh-Taylor problems which we shall refer to as $z$-models.  
These three $z$-models are nonlocal evolution equations for the interface parametrization $\boldsymbol z=(z^1,z^2,z^3)$ and the vortex sheet density on the interface $\boldsymbol\mu=(\mu_1,\mu_2)$.   The velocity is then obtained by Laplacian inversion via an approximation of the Birkhoff-Rott singular integral kernel.
Our $z$-models are presented in what we term low-order, medium-order, and high-order variants, corresponding to the amount of nonlocality each model retains from the original Euler system. 
We note that our initial derivation of the $z$-models are for a two-phase
flow with one material interface, but we shall also provide a generalization to the case of $n$ fluid interfaces with $n\ge 1$.

After deriving the $z$-models using our asymptotic method in non-locality,  we present a simple numerical discretization, which runs two orders of magnitude faster than traditional numerical methods for the Euler equations, and use it to study universality in RTI.
We begin with single-mode RTI and compute the evolution of the bubble and spike Froude numbers, dimensionless quantities measuring the ratio of inertial to gravitational effects on RT bubble and spike motion. 
Then we reproduce results from the classical ``rocket rig'' experiments of \citet{Re1984} and \citet{Yo1984}, on the growth rate of turbulent RT mixing layers, and compare against more recent models for mixing layer development of \citet{ClRi2004} and \citet{CaCo2006}. 
Due to the speed of our model, we are able to quickly compute emsemble-averaged quantities such as the density field, and we show that this closely matches the self-similar density distribution computed from a so-called Prandtl closure model. 
In addition, using the  standard mixing norms   introduced in   \citet{MaMePe2005}, we demonstrate that the ensemble-averaged density field (associated to many simulations initiated with randomly perturbed data) reproduces optimal mixing rates for a passive scalar, a property which has only been proven for specially chosen (shear-type)  transport velocities \citep{Th2012}.
Finally,  as we noted above, we generalize our $z$-model to allow for multiple fluid interfaces and use this to corroborate experimental evidence from \cite{DaJa2005} on the self-similarity of the species fraction profile in \emph{three}-phase fluid problems, which in turn shows that
the presence of multiple fluid layers has an important effect on the long-time growth of the mixing layer.

\subsection{Outline}
In Section \ref{sec::models}, we motivate and derive a class of interface models using a new type of asymptotic expansion in nonlocality. 
In Section \ref{sec::Euler}, we formulate the two-phase irrotational and incompressible Euler equations as a system of singular integral equations for the fluid interface parameterization and the vorticity measure on the interface. In Section \ref{sec::approximation}, we outline the difficulty of 
numerically simulating the boundary integral formulation of the Euler equations, and explain the need for simplified models for interface evolution 
in RTI. In Section \ref{sec::lower}, we compute the leading order terms in an asymptotic expansion of the Birkhoff-Rott kernel, and derive the 
so-called lower-order $z$-model, which is a fast modal model for interface evolution in RTI which allows for interface turnover. In Section 
\ref{sec::higher}, we introduce the medium- and higher-order $z$-models, which are modifications of the lower-order $z$-model which include 
increasing amounts of nonlocality from the original Euler system. Finally, in Section \ref{sec::vorticity}, we perform some basic analysis indicating 
that the modifications we have made in formulating the $z$-models are appropriate for the regime under consideration.

In Section \ref{sec::validation}, we describe a numerical discretization of the $z$-models, and validate our models against a variety of classical 
Rayleigh-Taylor test problems. In Section \ref{sec::numerics}, we describe our numerical method, and  demonstrate its large-scale 
convergence properties in Section \ref{sec::convergence}. 
We validate our model against experiments initialized with both single-mode initial data (Section \ref{sec::single-mode-data}) as well as random multi-mode initial data (Section \ref{sec::rocketrig}). 
In Section \ref{sec::mixing}, we compute ensemble-averaged quantities from many $z$-model runs using data with random fluctuations, and demonstrate that we are able to model viscous mixing layers with optimal mixing rates. 
In Section \ref{sec::mixing}, we use a so-called Prandtl closure model to produce a self-similar solution for the density profile, and show that it compares well with results from our $z$-model. 
In Section \ref{sec::stratified}, we use our $z$-model to study stratified flows containing unstable density interfaces, and demonstrate self-similarity in the ensemble-averaged species fraction profile.

\section{Interface models}\label{sec::models}

When the fluids are inviscid, incompressible, and irrotational in their respective bulks, the 3D Euler equations can be written as boundary integral equations involving only the interface position and the vortex sheet density on the interface. In three dimensions, these nonlocal integral equations were first derived and simulated by \citet{BaMeOr1984}, following the earlier derivation by \citet{Bi1962} and \citet{Ro1956} for  vortex sheets in 2D flows.
We consider two immiscible, inviscid,  incompressible, irrotational fluids occupying two open volumes $\Omega^+(t)$ and $\Omega^-(t)$ in $\mathbb{R}^3$, separated by a time-dependent material interface $\Gamma(t)$,  where
$t$ denotes time.  The  evolution of the fluid  is restricted to a time interval $0 \le t \le T$, and is modeled by the incompressible and irrotational Euler equations,
which can be written as
\begin{align} 
\left.
\begin{array}{r}
\rho ^\pm( \p_t\boldsymbol u^\pm +\boldsymbol u^\pm \bcdot \bnabla\boldsymbol u^\pm) + \bnabla p^\pm = - g \rho^\pm \boldsymbol e_3\\
\p_t \rho^\pm +\boldsymbol u^\pm \bcdot \bnabla \rho^\pm =0
\end{array}
\right\}
 \ \text{ in } \ \Omega^\pm(t) \,, \ \ \ 0< t\le T \,, 
 \label{euler00}
\end{align} 
with the constraints that
\begin{align*} 
\operatorname{div}\boldsymbol u^\pm =0 \ \text{ and } \ \operatorname{curl}\boldsymbol u^\pm =\boldsymbol 0 \ \text{ in } \ \Omega^\pm(t) \,, \ \ \ 0\le t\le T \,.
\end{align*} 
The initial interface $\Gamma(0)$ is specified and hence so too are the initial domains $\Omega^\pm(0)$, and initial conditions for
velocity and density are given:
\begin{align*} 
\boldsymbol u^\pm(\boldsymbol x,0) = \mathring{\boldsymbol u}^\pm(\boldsymbol x) \,, \ \ \  \rho^\pm (\boldsymbol x,0)= \mathring \rho^\pm (\boldsymbol x)   \ \text{ in } \ \Omega^\pm(0) \,.
\end{align*} 
Here, $\boldsymbol u^\pm$ denotes the velocity  and $p^\pm$ denotes the pressure in $\Omega^\pm(t)$, while $g$ represents the gravitational
acceleration, and the unit vector $\boldsymbol e_3=(0,0,1)$.   We assume constant  initial density functions 
$\rho^\pm$ in
$\Omega^\pm(0)$, which implies that $\rho^\pm$ remain constant functions in $\Omega^\pm(t)$.
Our focus is on flows in
which $\Gamma(t)$ denotes a vortex sheet, and hence we supplement the Euler equations with the following jump conditions on $\Gamma(t)$:
\begin{subequations} 
\label{jump}
\begin{align} 
\left.
\begin{array}{r}
\jump{p} =0 \\
\jump{\boldsymbol u \bcdot\boldsymbol n} =0 
\end{array} 
\right\} \ \text{ for } \ 0\le t\le T \,,
\label{jump-unp}
\end{align} 
where  $n$ denotes the  unit normal to $\Gamma(t)$, and $\jump{f} = f^+ - f^-$ on $\Gamma(t)$. In this manuscript we restrict our attention to flows where surface tension is negligible, but in general the condition $\jump{p}=0$ is replaced by the Young-Laplace law $\jump{p}=4\sigma H$, where $\sigma$ is the coefficient of surface tension and $H$ is the mean curvature of the interface $\Gamma(t)$.
We assume that $\rho^+ \neq \rho^-$ so that
\begin{align} 
\jump{\rho} \neq 0 \ \text{ on } \ \Gamma(t) \ \text{ for } \ 0\le t\le T \,. \label{jump-rho}
\end{align} 
Because of \eqref{jump-unp} and \eqref{jump-rho}, even if $\mathring{\boldsymbol u}^+ = \mathring{\boldsymbol u}^-$ on $\Gamma(0)$, the 
tangential component of velocity becomes discontinuous across $\Gamma(t)$ and we have that 
\begin{align} 
\jump{\boldsymbol u \bcdot \boldsymbol\tau_\alpha} \neq 0 \ \text{ on } \ \Gamma(t)  \ \text{ for } \ 0< t\le T \,, \ \ \alpha=1,2\,. \label{jump-utau}
\end{align} 
\end{subequations} 
Here,  $\boldsymbol\tau_1$ and $\boldsymbol\tau_2$ are  the unit tangent vectors to $\Gamma(t)$, chosen such that $(\boldsymbol\tau_1,\boldsymbol\tau_2,\boldsymbol n)$ form a right-handed orthonormal basis.   To complete the description of the dynamics,
the motion of the interface $\Gamma(t)$ is governed by the normal component of the fluid velocity.  Letting
$\mathcal{V}(\Gamma(t))$ denote the normal speed of the interface,
\begin{align} 
\mathcal{V} (\Gamma(t)) =\boldsymbol u \bcdot\boldsymbol n \,.  \label{Gamma-speed}
\end{align}

Let $(x^1,x^2,x^3)$ denote the standard Euclidean coordinates on $\mathbb{R}^3$, and let $(s^1,s^2)$ denote coordinates on $ \mathbb{R}^2  $, used to
parameterize $\Gamma(t)$.  Specifically, the time-dependent interface
$\Gamma(t)$ is parametrized by a smooth function $\boldsymbol z:\mathbb{R}^2 \times [0,T] \to\mathbb{R}^3$ and
\begin{align} 
\Gamma(t) = \{\boldsymbol z(s^1,s^2,t) \ : (s^1,s^2) \in \mathbb{R}^2  \,, \ t \in [0,T]\} \,, \,\, \boldsymbol z=(z^1,z^2,z^3) \,. \label{z-param}
\end{align} 
We use Latin indices for coordinates in Euclidean space, and Greek indices for coordinates on $\Gamma(t)$, and we will apply Einstein's summation convention without further comment. Euclidean space is endowed with the standard diagonal metric $\delta_{ij}$, and the induced metric on $\Gamma(t)$ is given by
$$h=h_{\alpha\beta}\mathrm{d}s^\alpha\otimes \mathrm{d}s^\beta\,,\qquad\qquad h_{\alpha\beta}=\partial_\alpha\boldsymbol z\bcdot\partial_\beta\boldsymbol z\,.$$
We set
$$|h|=\det h\,,\qquad\qquad \sqrt h=\sqrt{\det h}\,. $$
With respect to the parameterization \eqref{z-param}, the  time-dependent unit normal is given by
 $$\boldsymbol n= \frac{\partial_1\boldsymbol z\times\partial_2\boldsymbol z}{\sqrt h} \,. $$

As noted above, we assume that  the vorticity vanishes in the open sets $\Omega^+(t)$ and $\Omega^-(t)$,
 but on the material interface $\Gamma(t)$,  the tangential component of velocity experiences a jump discontinuity, resulting in a vorticity measure 
 $\boldsymbol\omega(s,t)$ concentrated 
 on the interface. The velocity in $\Omega^+(t)\cup \Omega^-(t)$ is computed from this vorticity measure using the Birkhoff-Rott integral
\begin{equation}\label{BR1}
	\boldsymbol u(\boldsymbol x,t)=\frac{1}{4\pi}\iint_{\mathbb{R}^2}\boldsymbol\omega(s,t)\times\frac{\boldsymbol x-\boldsymbol z(s,t)}{|\boldsymbol x-\boldsymbol z(s,t)|^3}\mathrm{d}s \,,
\end{equation}
where 
$$\boldsymbol\omega=\omega^\alpha\partial_\alpha\boldsymbol z$$ 
and $\omega^\alpha$, $\alpha=1,2$ denote the components of the vorticity measure with respect to the basis $\partial_\alpha\boldsymbol z$.   To be more precise, 
$\boldsymbol\omega(s,t)=\boldsymbol\omega(s^1,s^2,t)$
denotes the the vortex sheet {\it density} and is often referred to as the {\it amplitude of vorticity}  or simply the vorticity measure on $\Gamma(t)$.
When evaluated on $\Gamma(t)$ (or equivalently, along the parameterization $\boldsymbol z(s,t)$), the Birkhoff-Rott integral exists in the principal value sense, 
and gives the average velocity 
\begin{align} 
\bar{\boldsymbol u}=  \tfrac{1}{2} (\boldsymbol u^++\boldsymbol u^-) \circ\boldsymbol z \,, \label{baru-def}
\end{align} 
where the `$\circ$' denotes composition. In particular, by choosing the Lagrangian parameterization $\boldsymbol z(s,t)$ such that $\p_t \boldsymbol z(s,t) = \tfrac{1}{2} (\boldsymbol u^++\boldsymbol u^-) (\boldsymbol z(s,t),t)$ with $\boldsymbol z(s,0)=s^1\boldsymbol e_1+s^2\boldsymbol e_2$, we have that
\begin{equation}\label{BR2}
	\partial_t\boldsymbol z(s,t)=\bar{\boldsymbol u}(s,t):=\frac{1}{4\pi}\iint_{\mathbb{R}^2}  \boldsymbol\omega(s',t)\times\frac{\boldsymbol z(s,t)-\boldsymbol z(s',t)}{|\boldsymbol z(s,t)-\boldsymbol z(s',t)|^3}\mathrm{d}s' \,.
\end{equation}
where $s'$ is a dummy variable for integration. Let us remark that according to \eqref{Gamma-speed}, it is only the normal component of $\boldsymbol u$ that determines the shape of $\Gamma(t)$, and there exists an 
infinite-dimensional tangential reparameterization symmetry of the interface which does not change its shape.   This means that a different parameterization 
$\boldsymbol z(s,t)$ 
such that $\p_t \boldsymbol z \bcdot\boldsymbol n ( \boldsymbol z,t)  = \bar{\boldsymbol u}(\boldsymbol z,t) \bcdot  \boldsymbol n(\boldsymbol z,t)$ 
but 
 $\p_t \boldsymbol z \bcdot \boldsymbol\tau_\alpha(\boldsymbol z,t) \neq \bar{\boldsymbol u}(\boldsymbol z,t) \bcdot  \boldsymbol\tau_\alpha(\boldsymbol z,t)$ would provide the same shape for $\Gamma(t)$,
but would alter the distribution of particles along the interface.  For our purposes, the Lagrangian parameterization \eqref{BR2} is a convenient choice, as the resulting evolution equation \eqref{fe2} takes a particularly simple form. 

We may write the vorticity measure $\boldsymbol\omega$ on $\Gamma(t)$ in terms of the velocity jump
\begin{align} 
\boldsymbol w=(\boldsymbol u^+-\boldsymbol u^-)\circ\boldsymbol z \,, \label{u-jump}
\end{align} 
in the form
\begin{equation}\label{vort}
	\boldsymbol\omega=(\boldsymbol w\bcdot\partial_2\boldsymbol z)\partial_1\boldsymbol z-(\boldsymbol w\bcdot\partial_1\boldsymbol z)\partial_2\boldsymbol z \,,
\end{equation}
so that $\boldsymbol w\bcdot\partial_2\boldsymbol z= \boldsymbol\omega\bcdot\partial_1\boldsymbol z$ and $\boldsymbol w\bcdot\partial_1\boldsymbol z= - \boldsymbol\omega\bcdot\partial_2\boldsymbol z$.
Some of the difficulty of vortex methods stem from the computation of the Birkhoff-Rott velocity; specifically, 
 a straightforward numerical quadrature results in chaotic motion of the interface \citep{Kr1986}. Let us describe our method of evaluation. Equation \eqref{BR2} may be rewritten as
\begin{align} 
\bar{\boldsymbol u}(s,t)=\iint_{\mathbb{R}^2  }\boldsymbol\omega(s',t)\times\bnabla G(\boldsymbol z(s,t)-\boldsymbol z(s',t))\mathrm{d}s' \,, \label{u-bar}
\end{align} 
where $G(x)=(4\pi|x|)^{-1}$ is the fundamental solution of the Laplacian in $\mathbb{R}^3$. Note that \eqref{BR1} is defined everywhere in the bulk 
$\Omega^+(t)\cup\Omega^-(t)$ and has  arguments $(x^1,x^2,x^3,t)\in\mathbb{R}^3 \times [0,T]$, while \eqref{BR2} is well-defined
on the interface $\Gamma(t)$ and has arguments $(s^1,s^2,t)\in\mathbb{R}^2\times [0,t]$. To avoid the singularity in the integral \eqref{u-bar}, we replace 
$G$ by a regularized function
$$G^\epsilon(\boldsymbol x)=\frac{1}{4\pi\sqrt{\epsilon^2+|\boldsymbol x|^2}},$$
where $\epsilon>0$ and $G^\epsilon$ converges weakly to $G$ as $\epsilon\to 0$. This results in the regularized Birkhoff-Rott velocity
\begin{equation}\label{ubarepsilon}
	\bar{\boldsymbol u}^\epsilon(s,t)=\frac{1}{4\pi}\iint_{\mathbb{R}^2}\boldsymbol\omega(s',t)\times\frac{\boldsymbol z(s,t)-\boldsymbol z(s',t)}{\big(\epsilon^2+|\boldsymbol z(s,t)-\boldsymbol z(s',t)|^2)^{3/2}}\mathrm{d}s'.
\end{equation}
This particular regularization was chosen for algebraic simplicity and falls under the well-known umbrella of \emph{vortex blob methods} 
\citep[see, for example,][]{Kr1986b}.

 These are methods where the standard Green's function, satisfying $\Delta G=\delta$, is replaced with a regularized Green's function  satisfying 
$$\Delta G^\epsilon(\boldsymbol x)=\psi^\epsilon(\boldsymbol x):=\frac{1}{\epsilon^3}\psi\left(\frac{\boldsymbol x}{\epsilon}\right),$$
where $\psi$ is a smooth, nonnegative function with most of its mass near zero and $\int_{\mathbb{R}^3}\psi=1$. This kind of regularization has the effect of replacing a singular vorticity distribution $\boldsymbol\omega(\boldsymbol x)=\boldsymbol\omega_0\delta(\boldsymbol x-\boldsymbol x_0)$ with a smooth `vortex blob' of finite size $\boldsymbol\omega(\boldsymbol x)=\boldsymbol\omega_0\psi^\epsilon(\boldsymbol x-\boldsymbol x_0)$. For our choice of regularization, 
\begin{align} 
\psi(\boldsymbol x)=\frac{3}{4\pi}\frac{1}{(1+|\boldsymbol x|^2)^\frac{5}{2}}.
\label{kernel-desing}
\end{align} 
With this regularization employed, we may evaluate $\bar{\boldsymbol u}^\epsilon$ with any standard 2D quadrature method. This  allows us to obtain arbitrarily high-order spatial discretizations. 

The regularization applied here represents averaging over a length of size $\epsilon$, and has the effect of replacing an infinitely thin vortex sheet with a vortex sheet of finite width proportional to $\epsilon$. 
In particular, discontinuous quantities like the density $\rho$ are replaced by $\rho*\psi^\epsilon$, and the interface $\Gamma(t)$ represents the contour of volume fraction $\sfrac{1}{2}$, which has the effect of damping instability in the interface position at scales smaller than $\epsilon$. 
In the event of a mixing transition to a space-filling vorticity field, the interface still makes sense as the contour of volume fraction $\sfrac{1}{2}$.  
In the high-Reynolds number limit, the instability takes place simultaneously at all scales  down to the molecular, and is self-similar in the sense that the instability appears the same at every scale. 
Thus, by setting our regularization parameter $\epsilon$, we allow the vorticity to be space-filling at scales $\propto\epsilon$, which smears out the instability at length scales $\lambda\ll\epsilon$ while maintaining an accurate picture at scales $\lambda\gg\epsilon$.
For example, to simulate RTI in cloud formation as viewed from many kilometers away, one might set $\epsilon\sim 1\,\text{km}$ and achieve accurate results, despite the fact that the instability is taking place at all scales simultaneously. 

\subsection{The irrotational and incompressible  Euler equations}
\label{sec::Euler}
Thanks to the irrotationality of the two fluids, we may write the velocities $\boldsymbol u^\pm$ in $\Omega^\pm(t)$  in terms of velocity potentials,  $\boldsymbol u^\pm = \bnabla \varphi^\pm$,
where $\varphi^\pm$ are governed by Bernoulli's law,
\begin{equation}\label{bern}
	\begin{cases}
	\p_t\varphi^+ +\tfrac{1}{2}|\boldsymbol u^+|^2+gx^3=-\tfrac{p^+}{\rho^+}&\text{in }\Omega^+(t)\\
	\p_t\varphi^- +\tfrac{1}{2}|\boldsymbol u^-|^2+gx^3=-\tfrac{p^-}{\rho^-}&\text{in }\Omega^-(t)
\end{cases} \,,
\end{equation}
and where (we recall that) $p^\pm$ denotes the pressure functions, $\rho^\pm$ denotes the density functions,  and  $g$ denotes the gravitational acceleration. 
We assume that the two fluids have infinite extent, so that $\Omega^+(t)\cup \Gamma(t)\cup \Omega^-(t)=\mathbb{R}^3$. The jump conditions on $\Gamma(t)$
 associated to  \eqref{bern}  are given by \eqref{jump}.
 
 To determine how the potential jump $\varphi^+-\varphi^-$  varies along $\Gamma(t)$, we compose \eqref{bern} with the parameterization $\boldsymbol z$ and apply the chain rule:
\begin{align*}
	\partial_t((\varphi^+-\varphi^-)\circ\boldsymbol z)&=(\bnabla\varphi^+\circ\boldsymbol z-\bnabla\varphi^-\circ\boldsymbol z)\bcdot\partial_t\boldsymbol z+\p_t(\varphi^+ -\varphi^-)\circ\boldsymbol z=\boldsymbol w\bcdot\bar{\boldsymbol u}
	+\p_t(\varphi^+-\varphi^-)\circ\boldsymbol z \,, \\
	\partial_t((\varphi^++\varphi^-)\circ\boldsymbol z)&=(\bnabla\varphi^+\circ\boldsymbol z+\bnabla\varphi^-\circ\boldsymbol z)\bcdot\partial_t\boldsymbol z+\p_t(\varphi^+ +\varphi^-)\circ\boldsymbol z=2|\bar{\boldsymbol u}|^2
	+\p_t(\varphi^++\varphi^-)\circ\boldsymbol z \,. 
\end{align*}
Here,  $\boldsymbol w=\boldsymbol u^+-\boldsymbol u^-$ is the velocity jump on $\Gamma(t)$. Note that $\bar{\boldsymbol u}$, defined by \eqref{u-bar}, and $\boldsymbol w$  are well-defined on $\Gamma(t)$, and 
are thus functions of $(s^1,s^2)$ and $t$.
We introduce the {\it Atwood number}
$$A=\frac{\rho^+-\rho^-}{\rho^++\rho^-}\,.$$
By using \eqref{jump-unp}, we write
\begin{equation}\label{presdiff}
	\left(\frac{p^+}{\rho^+}-\frac{p^-}{\rho^-}\right)+A\left(\frac{p^+}{\rho^+}+\frac{p^-}{\rho^-}\right)=(A+1)\frac{p^+}{\rho^+}+(A-1)\frac{p^-}{\rho^-}=2\frac{p^+-p^-}{\rho^++\rho^-}=0.
\end{equation}
Using \eqref{bern}--\eqref{presdiff} and the identity
\begin{align*}
	|\boldsymbol u^\pm|^2=\left|\bar{\boldsymbol u}\pm\tfrac{1}{2}\boldsymbol w\right|=|\bar{\boldsymbol u}|^2\pm\bar{\boldsymbol u}\bcdot\boldsymbol w+\tfrac{1}{4}|\boldsymbol w|^2 \,,
\end{align*}
we find that along $\Gamma(t)$, 
\begin{subequations} 
\label{diffmean}
\begin{align}
\left(\frac{p^+}{\rho^+}-\frac{p^-}{\rho^-}\right) \circ \boldsymbol z &= \left(\p_t(\varphi^+-\varphi^-) +\tfrac{1}{2}\left(|\boldsymbol u^+|^2-|\boldsymbol u^-|^2\right)\right) \circ\boldsymbol z
	       =\partial_t((\varphi^+-\varphi^-)\circ\boldsymbol z) \,, \\
\left(\frac{p^+}{\rho^+}+\frac{p^-}{\rho^-} \right) \circ\boldsymbol z &=\left( \p_t(\varphi^++\varphi^-)+\tfrac{1}{2}\left(|\boldsymbol u^+|^2+|\boldsymbol u^-|^2\right)\right) \circ\boldsymbol z
      +2gz^3 \notag \\
      &=\partial_t((\varphi^+-\varphi^-)\circ\boldsymbol z)-|\bar{\boldsymbol u}|^2+\tfrac{1}{4}|\boldsymbol w|^2+2gz^3 \,.
\end{align}
\end{subequations} 
Substitution of \eqref{diffmean} into \eqref{presdiff} results in 
\begin{equation}\label{potEuler}
	\partial_t((\varphi^+-\varphi^-)\circ\boldsymbol z)+A\partial_t((\varphi^++\varphi^-)\circ\boldsymbol z)+A\left(-|\bar{\boldsymbol u}|^2+\tfrac{1}{4}|\boldsymbol w|^2+2gz^3\right)=0.
\end{equation}
It is convenient to introduce a ``rotated'' version of the vorticity components,
$$\mu_\alpha=\boldsymbol w\bcdot\partial_\alpha\boldsymbol z\,. \ \ \  (\alpha=1,2)\,, $$
Differentiating \eqref{potEuler} with respect to $s_\alpha$ then yields
\begin{equation}\label{Euler}
	\partial_t(\mu_\alpha+2A\bar{\boldsymbol u}\bcdot\p_\alpha\boldsymbol z)=A\partial_\alpha\big(|\bar{\boldsymbol u}|^2-\tfrac{1}{4}h^{\alpha\beta}\mu_\alpha\mu_\beta-2gz^3\big) \,,
\end{equation}
where $h^{\alpha\beta}$ is the $(\alpha,\beta)$ component of the inverse of the induced metric on $\Gamma(t)$:
$$h^{\alpha\beta}h_{\beta\gamma}={\delta^\alpha}_\gamma.$$
The coupled system of  equations \eqref{BR2} and \eqref{Euler} are written as
\begin{subequations} 
\label{full-euler}
\begin{align}
	\partial_t\boldsymbol z& =\bar{ \boldsymbol u} \,,  \label{fe1} \\
	\partial_t(\mu_\alpha+2A\bar{\boldsymbol u}\bcdot\partial_\alpha \boldsymbol z)& =A\partial_\alpha\big(|\bar{\boldsymbol u}|^2-\tfrac{1}{4}h^{\beta\gamma}\mu_\beta  \mu_\gamma-2gz^3\big) \,, \label{fe2}
\end{align}
\end{subequations} 
and are an equivalent form 
of the 3D irrotational and incompressible  Euler equations \eqref{euler00}, giving the interface position $\boldsymbol z=(z^1,z^2,z^3)$ and the components of velocity 
jump $\mu=(\mu_1,\mu_2)$ 
(or equivalently, the vorticity measure). 
This results in a form of the Euler equations which have 3 independent variables instead of the usual 4. 
The system \eqref{full-euler} is supplemented with the initial conditions $\boldsymbol z(s,0) =\mathring{\boldsymbol z}(s)$ and $\mu_\alpha(s,0)
= \mathring \mu_ \alpha (s)$. 

\subsection{The Need for Approximation}
\label{sec::approximation}
The numerical solution of the singular integral form of the Euler equations \eqref{full-euler} is difficult because of  the time-derivative term $\partial_t(2A\bar {\boldsymbol u}\bcdot\partial_\alpha\boldsymbol z)$ in \eqref{fe2}. Expanding this term  using \eqref{fe1} shows that
\begin{align*}
	\partial_t(2A\bar{\boldsymbol u}\bcdot\partial_\alpha\boldsymbol z)&=\frac{A}{2\pi}\iint_R\big(\mu_2(s)\partial_1\bar{\boldsymbol u}(s)-\mu_1(s)\partial_2\bar {\boldsymbol u}(s)\big)\times\frac{\boldsymbol z-\boldsymbol z(s)}{|\boldsymbol z-\boldsymbol z(s)|^3}\\
	&+\big(\partial_t\mu_2(s)\partial_1\boldsymbol z(s)-\partial_t\mu_1(s)\partial_2\boldsymbol z(s)\big)\times\frac{\boldsymbol z-\boldsymbol z(s)}{|\boldsymbol z-\boldsymbol z(s)|^3}\\
	&-\big(\mu_2(s)\partial_1\boldsymbol z(s)-\mu_1(s)\partial_2\boldsymbol z(s)\big)\times\big(\boldsymbol z-\boldsymbol z(s)\big)\frac{3(\bar{\boldsymbol u}-\bar{\boldsymbol u}(s))\bcdot(\boldsymbol z-\boldsymbol z(s))}{|\boldsymbol z-\boldsymbol z(s)|^5}\mathrm{d}s \,.
\end{align*}
where we have not written the time-dependence in the integrand.
Hence we see that \eqref{full-euler} takes the form of a system of  nonlinear and nonlocal integral equations for the time-derivatives 
$\p_t(z^1,z^2,z^2,\mu_1,\mu_2)$. 
In addition to the difficulty of solving \eqref{full-euler} caused by the term $\partial_t(2A\bar{\boldsymbol u}\bcdot\p_\alpha\boldsymbol z)$, yet another challenge stems from
fact that the Birkhoff-Rott integral can be time-consuming to evaluate: a naive implementation takes $O(N^2)$ operations for an $N$-point discretization
of $\Gamma(t)$, although this can be  improved greatly by the use of  fast multipole methods.  We are not aware of any successful attempts to directly 
simulate the singular integral form of the Euler equations \eqref{full-euler}, despite numerous simulations  of these equations in two space dimensions 
and with axisymmetry. The models which we derive in the following sections make the appropriate reductions of the full incompressible and
irrotational Euler equations so as to retain high accuracy in the interface position while avoiding the considerable numerical challenges that
we have explained above.

In two dimensions, \cite{GrSh2017} remedied the computational difficulty of the Birkhoff-Rott velocity by introducing an approximate velocity which can be computed as a Hilbert transform, requiring only $O(N\log N)$ evaluations:
\begin{align}
	\widetilde{\boldsymbol u}=\frac{H\omega}{2|\p_{1}\boldsymbol z|}\boldsymbol n,&& \widehat{Hf}(k)=-\mathrm{i}\,\text{sgn}(k)\hat f(k).  \label{approx-u2d}
\end{align}
The system \eqref{approx-u2d} is the approximation which results from taking a limit of ``small nonlocality'' (explained in detail in the sequel), which is 
accurate when the interface is not too curved; in particular, it was shown in \cite{GrSh2017} that
\begin{equation}
	\max_{s \in \mathbb{R}  }|\bar{\boldsymbol u}(s,t)-\widetilde{\boldsymbol u}(s,t)|\leq\frac{2}{\pi}\sqrt{3K} \int_{\Gamma(t)} \omega(s,t)^2  \sqrt{h} \mathrm{d}s\,,
\end{equation}
where $K$ is the maximum curvature of the interface. Inserting the approximate velocity \eqref{approx-u2d} into the 2D version of \eqref{full-euler} provides
a simple set of evolution equations for $\boldsymbol z$ and $\omega$, called the 2D \emph{lower-order $z$-model}. It was subsequently verified in \cite{CaDeFrGoRaReSh2020} that the lower-order $z$-model agrees well with experimental data of Rayleigh-Taylor problems, forming a sort of ``envelope'' for the interface roll-up.

\subsection{The 3D Lower-Order $z$-Model}
\label{sec::lower}
The aim of this section is to introduce a 3D generalization of the  2D lower-order $z$-model of \citet{GrSh2017}, as well as two more accurate models (with greater
nonlocality), which we shall refer to as the ``medium-order'' and ``higher-order'' $z$-models. While already useful for 2D simulations (in respect to the speed-up over
traditional Euler solvers), the need for such a simplification is even greater for 3D simulations.

\begin{figure} 
	\centering
	\includegraphics[width=6in]{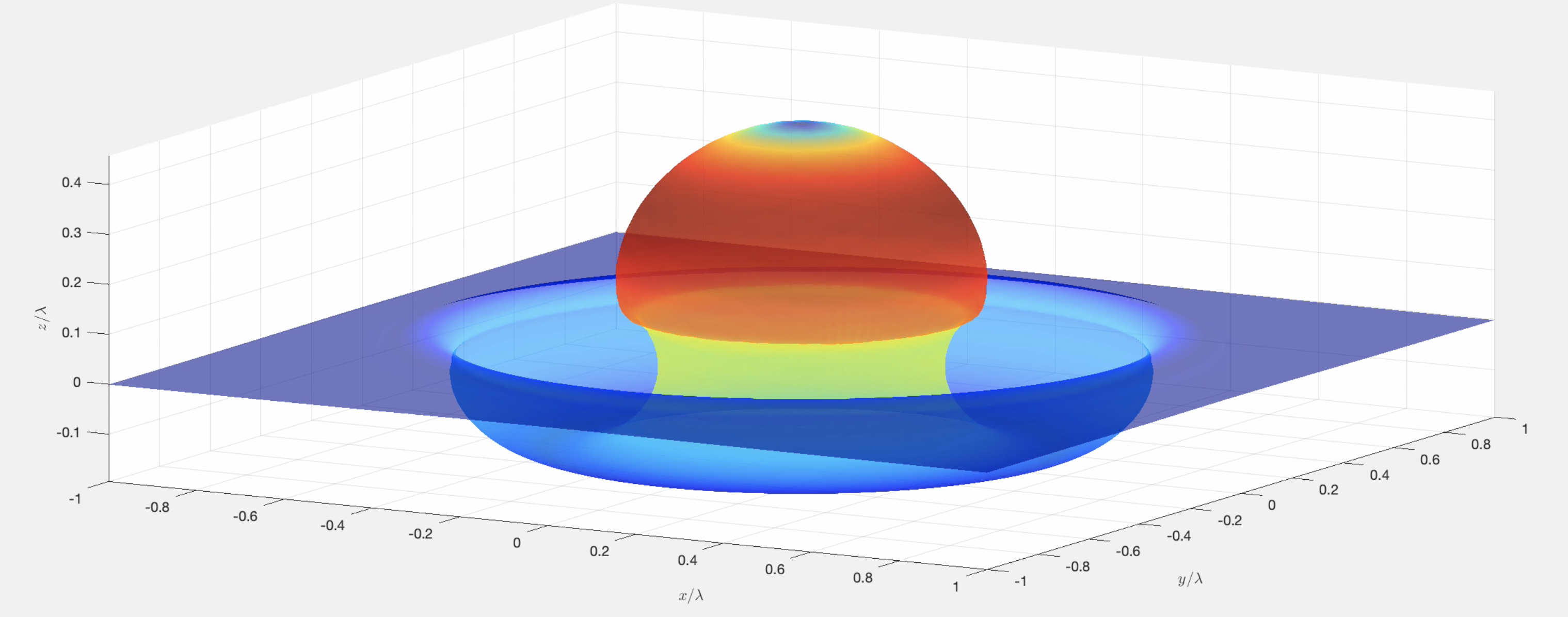}
	\caption{\footnotesize{Evolution of the lower-order $z$-model, starting from a Gaussian perturbation of a flat interface, run on a $192\times192$ mesh in 18s. }}
	\label{bubble_lo}
\end{figure}

For the remainder of this section, we omit writing the explicit dependence on time $t$.
The derivation begins with the observation that the dominant contribution to the Birkhoff-Rott velocity $\bar{\boldsymbol u}(s)$ in \eqref{BR1} arises from the
singular integrand in a small neighborhood of $s$.    
Expanding $\boldsymbol z(s')$ in a small neighborhood of $s$ yields
\begin{align} 
\boldsymbol z(s') = \boldsymbol z(s) + \p_ \alpha\boldsymbol z(s) ({s'}^ \alpha - s^ \alpha ) + \tfrac{1}{2} \p_ {\alpha \beta}\boldsymbol z(s) ({s'}^ \alpha - s^ \alpha ) ({s'}^ \beta - s^ \beta )  + O(|s-s'|^3) \,,
\label{temp0}
\end{align} 
and the expansion of $\omega(s')$ about $\omega(s)$ is given by
\begin{align} 
\boldsymbol\omega(s') & = \mu_2(s') \p_1\boldsymbol z(s') - \mu_1(s') \p_2\boldsymbol z(s')  \notag \\
& = \mu_2(s') \p_1\boldsymbol z(s) - \mu_1(s') \p_2\boldsymbol z(s) \notag \\
&+  \mu_2(s')  \p_{1 \alpha }\boldsymbol z(s) ({s'}^\alpha -s^ \alpha ) - \mu_1(s')  \p_{2 \alpha }\boldsymbol z(s)  ({s'}^ \alpha - s ^ \alpha )
+O(|s-s'|^2) \,. \label{temp1}
\end{align}
The first-order expansion in \eqref{temp0} is useful in regions where the interface curvature is small, which is a valid assumption prior to interface roll-up or near bubble and spike tips. See section \ref{sec::vorticity} for precise estimates of the error between the approximate and exact velocities, and compare figures \ref{bubble_lo} and \ref{bubble_hi} to see the effect of this low-curvature assumption.
In order to expand $|\boldsymbol z(s)-\boldsymbol z(s')|^{-1}$ about the singular point $s=s'$,  we expand $\boldsymbol z(s)-\boldsymbol z(s')$ about $s=s'$, factor the linear term of this expansion, and use
the series expansion for $ 1/(1-\xi) $ about $ \xi=0$; in particular, we have that
\begin{align}
	\frac{1}{|\boldsymbol z(s)-\boldsymbol z(s')|^3}&=\frac{1}{|\partial_\alpha\boldsymbol z(s)(s'^\alpha-s^\alpha)|^3}\left(\frac{|\boldsymbol z(s)-\boldsymbol z(s')|}{|\partial_\alpha\boldsymbol z(s)(s'-s)|}\right)^{-3} \notag \\
	&=\frac{1}{|\partial_\alpha\boldsymbol z(s)(s'^\alpha-s^\alpha)|^3}
	\left(1+\frac{1}{2}\frac{\partial_{\alpha\beta}\boldsymbol z(s)(s'^\alpha-s^\alpha)(s'^\beta-s^\beta)}{|\partial_\alpha\boldsymbol z(s)(s'^\alpha-s^\alpha)|}+\cdots\right)^{-3}\notag \\
	&=\frac{1}{|\partial_\alpha\boldsymbol z(s)(s'^\alpha-s^\alpha)|^3}\left(1-\frac{3}{2}\frac{\partial_{\alpha\beta}\boldsymbol z(s)(s'^\alpha-s^\alpha)(s'^\beta-s^\beta)}{|\partial_\alpha\boldsymbol z(s)(s'^\alpha-s^\alpha)|}+\cdots\right)
	\label{temp3}
\end{align}
The expansions \eqref{temp0}--\eqref{temp3}, together with a somewhat tedious computation, show that the Birkhoff-Rott integrand 
$\omega(s')\times(\boldsymbol z(s)-\boldsymbol z(s'))/|\boldsymbol z(s)-\boldsymbol z(s')|^3$ has the following expansion about $s=s'$:
\begin{align}
&\frac{(s^\alpha-s'^\alpha)}{\big(h_{\alpha\beta}(s)(s^\alpha-s'^\alpha)(s^\beta-s'^\beta)\big)^\frac{3}{2}}
\Big( \mu_\alpha(s') 
+(s^\beta-s'^\beta) \big( \mu_\alpha(s') \p_\beta\sqrt{h}-  \tfrac{1}{2}\partial_{\alpha\beta}\boldsymbol z\bcdot\partial_\gamma z)h^{\gamma\delta}(s)\mu_\delta(s') \big)
\Big) \boldsymbol n \notag \\
&
\qquad 
+\frac{(s^\alpha-s'^\alpha)(s^\beta-s'^\beta)}{\big(h_{\alpha\beta}(s)(s^\alpha-s'^\alpha)(s^\beta-s'^\beta)\big)^\frac{3}{2}}\Big(\mu_\alpha(s')
\sqrt h\partial_\beta\boldsymbol n
-\tfrac{1}{2}(\mu_2(s')\partial_1\boldsymbol z-\mu_1(s')\partial_2\boldsymbol z)\times\boldsymbol n(\p_{\alpha\beta}\boldsymbol z\bcdot\boldsymbol n)\Big) \notag \\
&
\qquad
-\frac{3}{2}\frac{\mu_\alpha(s')(s^\alpha-s'^\alpha)(s^\beta-s'^\beta)(s^\gamma-s'^\gamma)(s^\delta-s'^\delta)(\partial_{\alpha\beta}\boldsymbol z\bcdot\partial_\gamma\boldsymbol z)\partial_\delta\boldsymbol z}{\big(h_{\alpha\beta}(s)(s^\alpha-s'^\alpha)(s^\beta-s'^\beta)\big)^\frac{5}{2}}\boldsymbol n+O(1) \,.
	\label{series0}
\end{align}
Isolating the dominant contribution, we thus have that
\begin{align}
\boldsymbol\omega(s')\times\frac{\boldsymbol z(s)-\boldsymbol z(s')}{|z(s)-z(s')|^3}
&=\frac{\mu_1(s')(s^1-{s'}^1)+\mu_2(s')(s^2-{s'}^2)}{\big(h_{\alpha\beta}(s)({s'}^\alpha-s^\alpha)({s'}^\beta-s^\beta)\big)^{3/2}}
\sqrt{h(s)}\boldsymbol n(s)+O(|s-s'|^{-1}) \,. \label{series1}
\end{align}
We can then integrate the leading order term  in \eqref{series1}
and obtain an approximate (average) velocity along the interface;  however,  the resulting velocity  field  is difficult to write in terms of simple Fourier multipliers
(which is our objective), and therefore maintains the computational expense of the original Birkhoff-Rott velocity.   
In order to derive a velocity field which takes advantage of the computational efficiency of Fourier multipliers, we must assume that the metric $h$ is isotropic, by which we mean that the components $h_{ \alpha \beta}$ of the
metric satisfy
 $$ h_{11} = h_{22}\,, \ \ \  h_{12} =h_{21} =0 \,, $$
 in which case
\begin{align} 
\frac{\sqrt{h(s)}\mu_\alpha(s')(s^\alpha-{s'}^\alpha)}{\big(h_{\alpha\beta}(s')({s'}^\alpha-s^\alpha)({s'}^\beta-s^\beta)\big)^{{\frac{3}{2}} }}\boldsymbol n(s)
=\frac{\mu_\alpha(s')(s^\alpha-{s'}^\alpha)}{|h(s)|}\boldsymbol n(s)\,. \label{temp2}
\end{align} 
Replacing the original integrand in the Birkhoff-Rott integral \eqref{BR2} with the first-order term \eqref{temp2} in its expansion
 results in an approximate velocity $\widetilde{\boldsymbol u}$,  computable in terms of \emph{Riesz transforms} $R^ \alpha $:
\begin{align}\label{utilde}
	\widetilde{\boldsymbol u}=\frac{R^\alpha\mu_\alpha}{2|h|}\boldsymbol n,&&R^\alpha f(s)=\frac{1}{2\pi}\iint_{\mathbb{R}^2}\frac{(s^\alpha-s'^\alpha)f(s')}{|s-s'|^3}\mathrm{d}s' \,, \ \alpha =1,2 \,.
\end{align}
Riesz transforms are Fourier multipliers: $\widehat{R^\alpha f}(k)=-ik^\alpha\hat f(k)/|k|$, where $\hat f(k)$ denotes the
Fourier transform of $f(x)$.   These operators are the multi-dimensional generalization of the Hilbert transform used in the definition of the lower-order 2D
$z$-model \eqref{approx-u2d}.

The first term of the Laurent series for 
$\boldsymbol\omega(s')\times(\boldsymbol z(s)-\boldsymbol z(s'))/|z(s)-z(s')|^3$ is given by \eqref{series1} with an error which is $O(|s-s'|^{-1})$ as $s' \to s$.   
The $O(|s-s'|^{-1})$ term in this expansion is what remains in  \eqref{series0}, and upon integration in $s'$,  produces a leading-order estimate  for the error or 
{\it remainder} $\mathcal{R} $ between the lower order velocity $\widetilde{\boldsymbol u}$ and the full 
Birkhoff-Rott velocity $\bar{\boldsymbol u}$. Utilizing our isotropy assumption shows that
\begin{align}
	\mathcal R=\frac{\partial_\beta(\sqrt h\boldsymbol n)R^\beta R^\alpha\mu_\alpha}{|h|^{3/2}}
	&-\frac{1}{2}\frac{h^{\gamma\delta}(\partial_{\alpha\beta}\boldsymbol z\bcdot\partial_\gamma\boldsymbol z)R^\beta R^\alpha\mu_\delta}{|h|^{3/2}}\boldsymbol n \notag\\
	&-\frac{1}{2}\frac{(\boldsymbol n\bcdot \p_{\alpha\beta}\boldsymbol z)\big((\partial_1\boldsymbol z\times\boldsymbol n)R^\alpha R^\beta\mu_2-(\partial_2\boldsymbol z\times\boldsymbol n)R^\alpha R^\beta\mu_1\big)}{|h|^{3/2}}\nonumber\\
	&-\frac{3}{2}\frac{(R^\alpha R^\beta R^\gamma R^\delta\mu_\alpha)(\partial_\beta\boldsymbol z\bcdot\partial_{\gamma\delta}\boldsymbol z)}{|h|^{5/2}}\boldsymbol n \,.
	\label{eqn::error}
\end{align}
Using the fact that the Riesz transforms have unit norm as bounded operators on $L^2$, H\"{o}lder's inequality shows that
\begin{equation}
	\label{eqn::errorestimate}
	\left(\iint_{\mathbb{R}^2} \mathcal{R} (s)^2ds\right)^{{\frac{1}{2}} }\leq C\left(\iint_{\mathbb{R}^2}|\mu(s)|_h^2 \mathrm{d}s\right)^{\frac{1}{2}}\sum_{\alpha,\beta=1}^2\sup_{s\in \mathbb{T}^2}\frac{|\p_{\alpha\beta}\boldsymbol z|}{1+|h|},
\end{equation}
where $|\mu|_h^2=h_{\alpha\beta}\mu^\alpha\mu^\beta$ is the magnitude of $\mu$ with respect to the metric $h$ on $\Gamma(t)$.
 In more concise notation, 
\begin{align} 
 \|\mathcal{R} \|_{L^2}\leq\|\mu\|_{L^2} \left\| \frac{D^2z}{|h|^{\frac{3}{2}}} \right\|_{L^\infty} \,.  \label{crude-bound}
\end{align} 
 Since the mean curvature vector on $\Gamma(t)$ is defined as
 $$
 \kappa(s,t) \boldsymbol n= \frac{h^{ \alpha \beta }}{ \sqrt h} \p_{ \alpha \beta }\boldsymbol z \bcdot (\p_1 z \times \p_2\boldsymbol z) \,,
 $$
the bound \eqref{crude-bound} becomes small when the mean curvature is very small, as $\|\mu\|_{L^2}$ remains bounded under this scenario.

Inserting $\widetilde{\boldsymbol u}$ into the Euler equations \eqref{Euler} in place of the Birkhoff-Rott velocity gets rid of the problematic time-derivative term 
$(\bar{\boldsymbol u}\bcdot\partial_\alpha\boldsymbol z)_t$, as $\widetilde{\boldsymbol u}$ is normal to $\Gamma(t)$, and so  $\widetilde{\boldsymbol u} \bcdot \partial_\alpha\boldsymbol z=0$. This yields the following system of equations:
\begin{subequations} 
\label{LO}
\begin{alignat}{2}
	\partial_t\boldsymbol z&=\widetilde{\boldsymbol u} \qquad && 0< t\le T \,,\label{LO_Z} \\ 
	\partial_t\mu_\alpha&=A\partial_\alpha\big(|\widetilde{\boldsymbol u}|^2-\tfrac{1}{4}h^{\beta\gamma}\mu_\beta \mu_\gamma-2gz^3\big)
	\qquad && 0< t\le T  \,,\label{LO_W} \\
	\boldsymbol z(s,0) & =\mathring{\boldsymbol z}(s) \,, \ \ \ \mu_\alpha(s,0) = \mathring \mu_\alpha(s)   \,,
\end{alignat}
\end{subequations} 
where $\mathring{\boldsymbol z}(s)$  and  $ \mathring \mu_\alpha(s)$ denote the initial conditions for the interface parameterization and vorticity, respectively.
We refer to \eqref{LO} as the \emph{lower-order $z$-model}.
It is an extremely efficient model, which tracks the interface accurately until about the time of turnover, after which it forms a sort of ``envelope'' for the roll-up in the near-symmetric regime (see Figure \ref{bubble_lo}). This envelope behavior accurately captures the large-scale behavior of RTI for interfaces with
 small curvature and small anisotropy, 
 although our medium- and higher-order $z$-models (which are introduced below) are better suited for capturing the small-scale structure of roll-up.  
In particular, the lower-order $z$-model suppresses interactions between interface points which are close in space but distant in interface variables, i.e. where $|\boldsymbol z(s)-\boldsymbol z(s')|\ll|s-s'|$. This occurs precisely when the interface folds over or rolls up, which is why the model produces poorer results after interfce turnover. At a point of high curvature, where the interface folds over itself, the full Birkhoff-Rott velocity correctly gives a large velocity and causes the interface to roll-up further, whereas the lower-order velocity remains small.

We note that a convenient feature of the lower-order $z$-model \eqref{LO} is that the velocity $\widetilde{\boldsymbol u}$ is given by a Fourier multiplier.  As such, the numerical
implementation of \eqref{LO} can take advantage of the Fast Fourier Transform which is both fast and easy to implement, and is one of advantages of this
model in comparison to the higher-order models to be introduced next. The computational speed of the  lower-order $z$-model creates an efficient and  effective  tool for determining large-scale features of RTI structures, such
 as  interface growth rates and bubble and spike positions.  On average, 
the lower-order $z$-model runs  about 15,000 times  faster than standard numerical schemes for 3D hydrodynamics \citep{Reisner}.

\subsection{The 3D Medium-Order and Higher-Order $z$-Models}
\label{sec::higher}
\begin{figure} 
	\centering
	\includegraphics[width=6in]{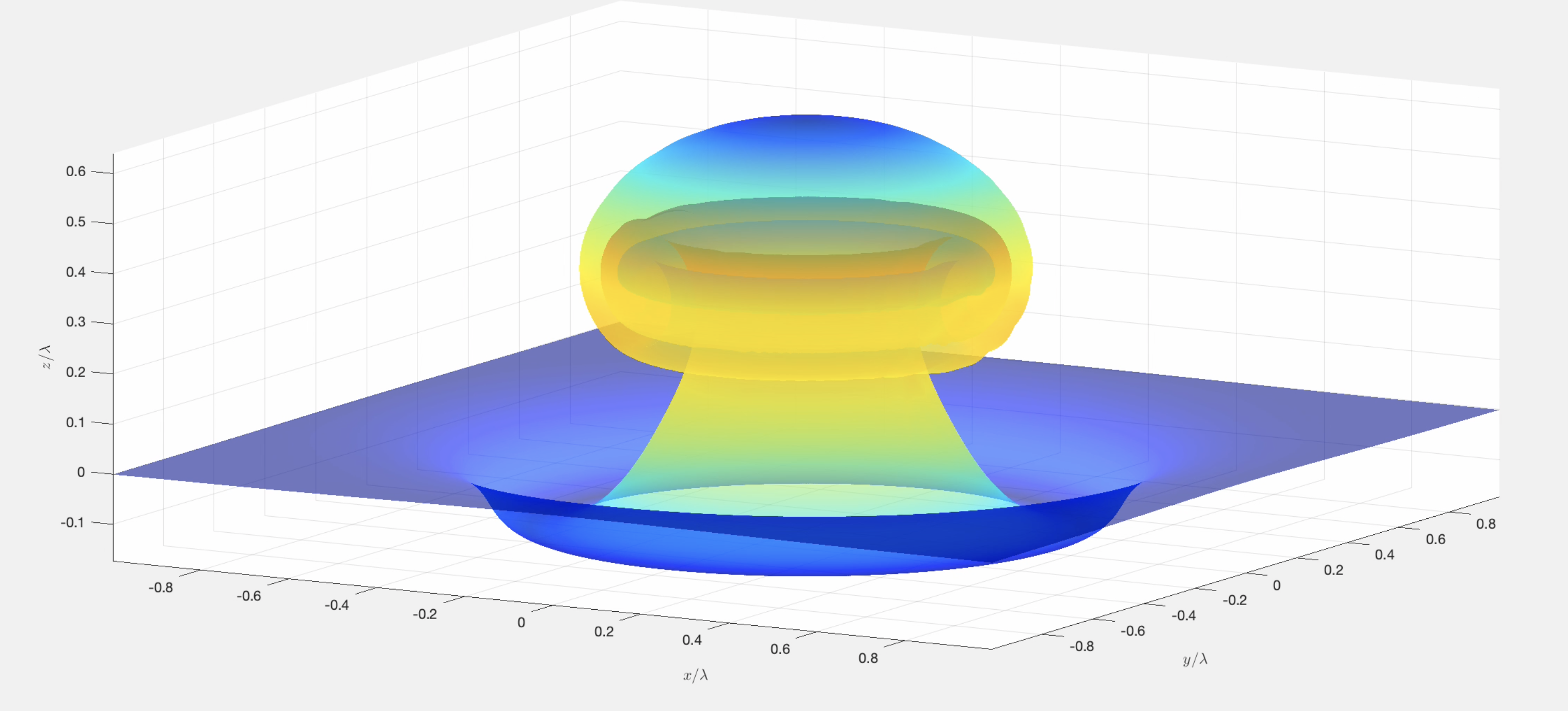}
	\caption{{\footnotesize Evolution of the higher-order $z$-model, starting from a Gaussian perturbation of a flat interface, run on a $192\times192$ mesh in 420s. Note that the lower-order $z$-model (Figure \ref{bubble_lo}) traces a sort of `envelope' for the higher-order $z$-model.  }}
	\label{bubble_hi}
\end{figure}

The isotropy assumption  made in the derivation of the lower-order  $z$-model is too restrictive for a large class of initial data.   As such,
we turn our attention to the medium-order $z$-model \eqref{MO} and higher-order $z$-model \eqref{HO}.  
The medium-order $z$-model  is obtained by replacing the localized velocity \eqref{utilde} with the regularized Birkhoff-Rott velocity \eqref{ubarepsilon} {\it only}
in  the $z$-evolution equation in \eqref{LO_Z} while keeping the vorticity equation \eqref{LO_W} unchanged:
\begin{subequations} 
\label{MO}
\begin{alignat}{2}
	\partial_t\boldsymbol z&=\bar{\boldsymbol u}^\epsilon \qquad && 0< t\le T \,,\label{MO_Z} \\ 
	\partial_t\mu_\alpha&=A\partial_\alpha\big(|\widetilde{\boldsymbol u}|^2-\tfrac{1}{4}h^{\alpha\beta}\mu_\alpha \mu_\beta-2gz^3\big)
	\qquad && 0< t\le T  \,,\label{MO_W} \\
	\boldsymbol z(s,0) & =\mathring{\boldsymbol z}(s) \,, \ \ \ \mu_\alpha(s,0) = \mathring \mu_\alpha(s)   \,,
\end{alignat}
\end{subequations} 
For the higher-order $z$-model, we replace the localized velocity with the regularized Birkhoff-Rott velocity in both equations of \eqref{LO}
to obtain
\begin{subequations} 
\label{HO}
\begin{alignat}{2}
	\partial_t\boldsymbol z&=\bar{\boldsymbol u}^\epsilon \qquad && 0< t\le T \,,\label{HO_Z} \\ 
	\partial_t\mu_\alpha&=A\partial_\alpha\big(|\bar{\boldsymbol u}^\epsilon|^2-\tfrac{1}{4}h^{\alpha\beta}\mu_\alpha \mu_\beta-2gz^3\big)
	\qquad && 0< t\le T  \,,\label{HO_W} \\
	\boldsymbol z(s,0) & =\mathring{\boldsymbol z}(s) \,, \ \ \ \mu_\alpha(s,0) = \mathring \mu_\alpha(s)   \,,
\end{alignat}
\end{subequations} 
The medium- and higher-order $z$-models are able to capture the fine-scale structures of RTI, including the Kelvin-Helmholtz roll-up regions (see 
Figure \ref{bubble_hi}), but are more costly to simulate numerically than the lower-order $z$-model. Nevertheless, the medium- and higher-order $z$-models run $600$ times faster than standard numerical methods for 3D gas dynamics codes \citep{Reisner}.

The evolution equations for the higher-order $z$-model differ from the integral form of the Euler equations by only one term.  In particular, a comparison of
the right-hand sides of
\eqref{fe2} with  \eqref{HO_W} shows that the higher-order $z$-model does not have the problematic nonlinear and nonlocal
time-derivative term $2A\partial_t(\bar{\boldsymbol u}\bcdot\p_\alpha\boldsymbol z)$.   As we will explain in Section \ref{sec::vorticity}, the reason that the higher-order $z$-model
maintains accuracy is that this time-derivative term tends to be small.   We note that for the numerical simulations we consider in this work, results from
the  medium-order $z$-model and the higher-order $z$-model are essentially indistinguishable (compare the first and second rows in Figure \ref{onemode_comparison}), but the higher-order $z$-model is more numerically stable.  
For this reason,  we have chosen not to report results from the medium-order $z$-model (with the exception of the second row in Figure \ref{onemode_comparison}).

\begin{figure} 
	\centering
	\includegraphics[width=6in]{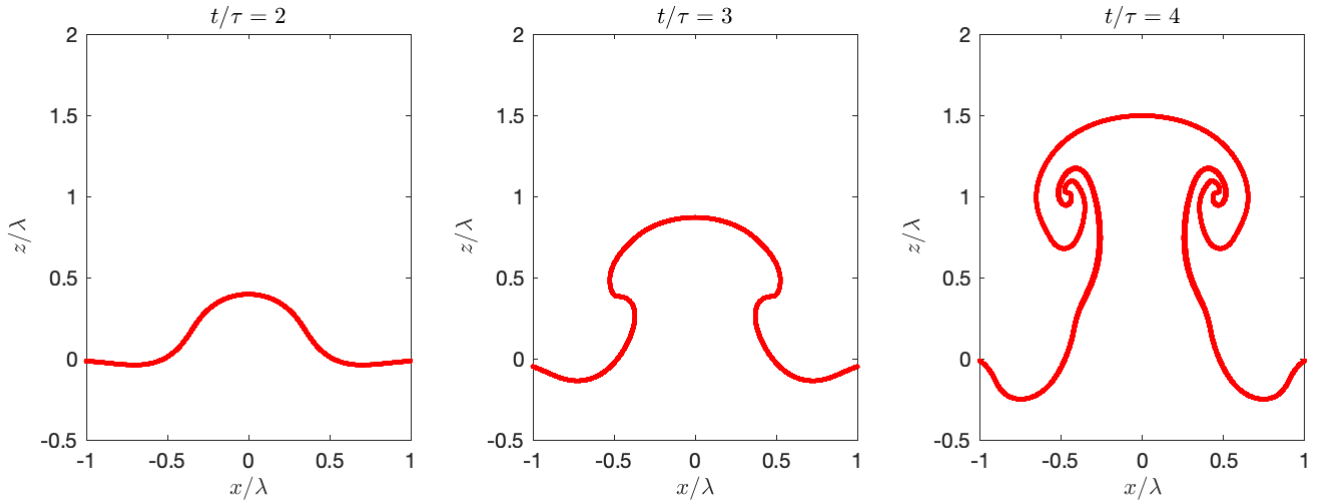}
	\includegraphics[width=6in]{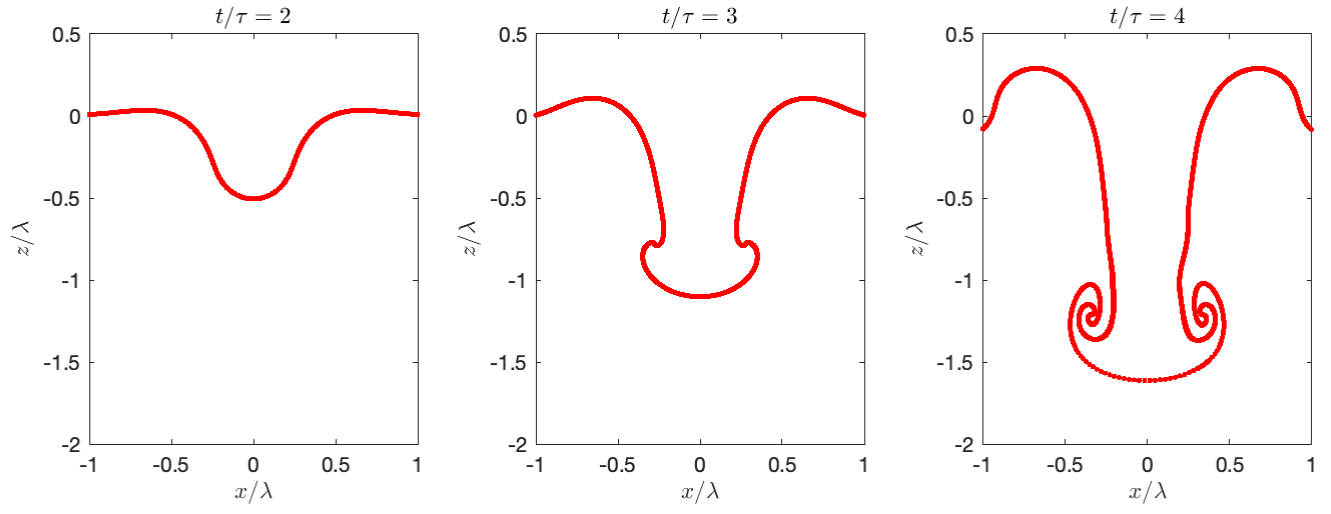}
	\caption{{\footnotesize Cross-sections of the 3D higher-order $z$-model, run on a $64\times 64$ grid, with Atwood number $0.7$ and a runtime of 50s.}}
	\label{bubblespike}
\end{figure}

An important feature of Rayleigh-Taylor instability is the differing behavior of `bubbles' of light fluid rising into the heavy fluid and `spikes' of heavy fluid falling into the light fluid. Our model accurately captures this differing behavior with the spikes being thinner and slightly longer than the bubbles. Figure \ref{bubblespike} shows the evolution of the initial conditions
\begin{equation}\label{gaussian}
	z^1=s^1,\qquad z^2=s^2,\qquad z^3=\pm0.05e^{-9|s|^2},\qquad \mu_1=\mu_2=0
\end{equation}
at two, three, and four characteristic times. 

\subsection{Dynamics of Vorticity}
\label{sec::vorticity}

\begin{figure} 
	\centering
	\includegraphics[width=6in]{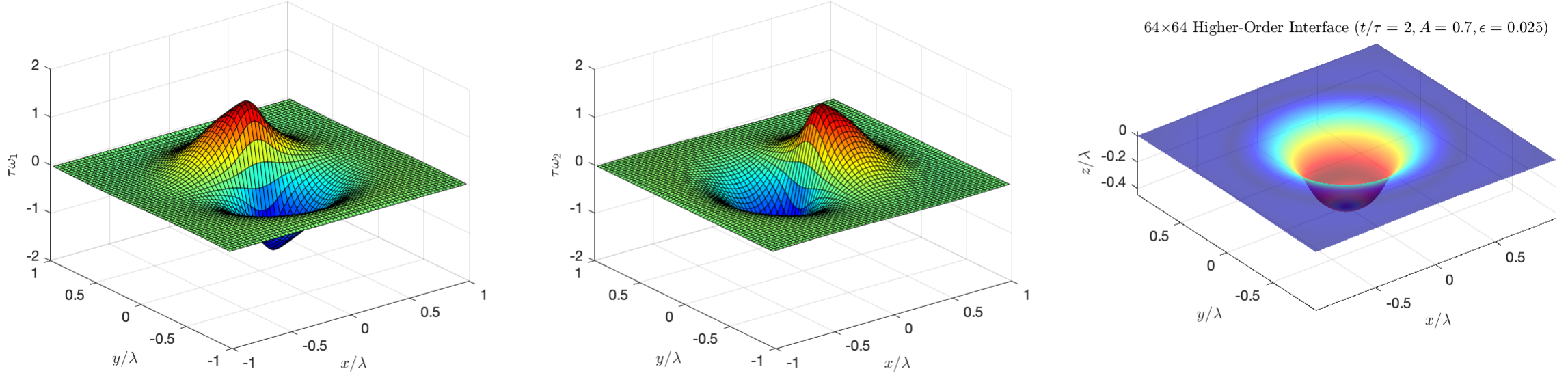}
	\includegraphics[width=6in]{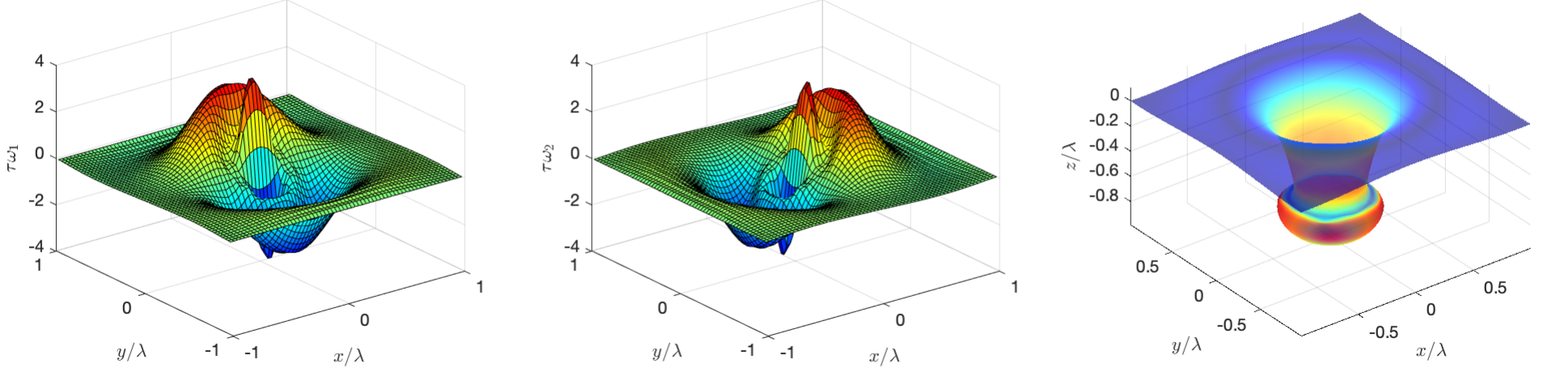}
	\caption{{\footnotesize Nondimensionalized components of vorticity for the higher-order $z$-model, compared with the full interface. Color on the interface indicates magnitude of vorticity (red is higher, blue is lower).}}
	\label{vorticity-components}
\end{figure}

\begin{figure}
	\centering
	\includegraphics[width=6in]{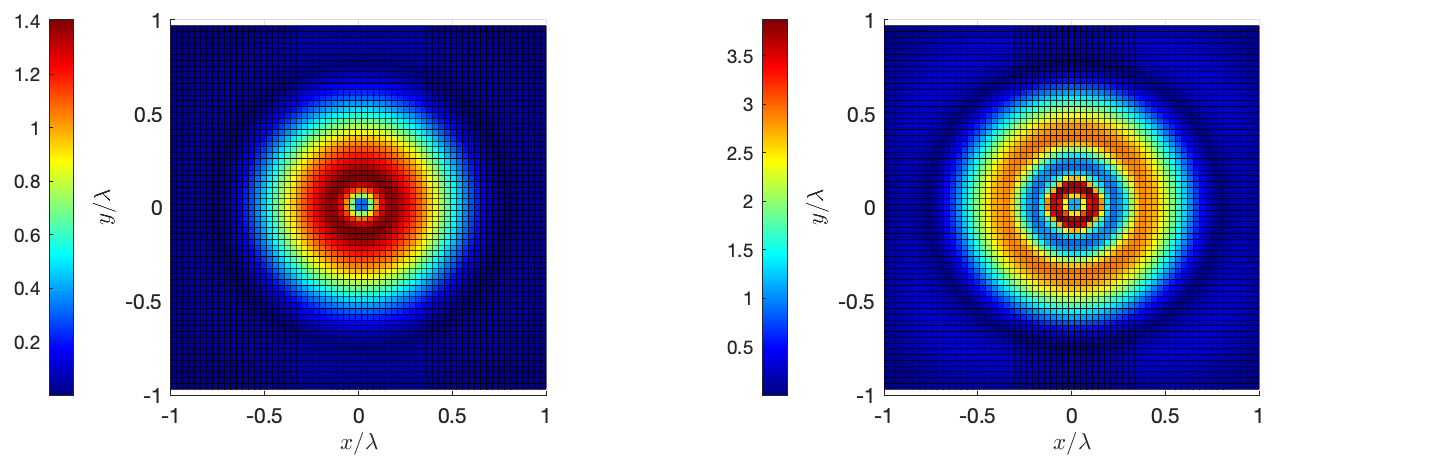}
	\caption{{\footnotesize Heatmap of the nondimensionalized magnitude of vorticity for the two interfaces shown in Figure \ref{vorticity-components}).}}
	\label{vorticity-magnitude}
\end{figure}

The relationship between the components of vorticity $\omega^\alpha$ and the interface roll-up is quite straightforward. Each `roll' of the interface corresponds to 
a sequence of successively stronger concentric `ridges' in the magnitude of vorticity, and the vector field $(\omega^1,\omega^2)$ circulates around these ridges. 
Figure \ref{vorticity-components} shows a simulation of the higher-order $z$-model \eqref{HO} in which a single ring of vorticity (top row, pre-turnover) splits 
into two rings of vorticity, the inner ring stronger than the outer (bottom row, post-turnover). 
Figure \ref{vorticity-magnitude} shows the corresponding magnitude of vorticity for the interfaces in   Figure \ref{vorticity-components}. These spikes in the vorticity are essential to the roll-up of the interface $\Gamma(t)$, but they are also a source of numerical instability. To mitigate this instability, we introduce a  smooth, vorticity-scaled,  nonlinear artificial viscosity, outlined in Section \ref{sec::numerics}.

\begin{figure}
	\centering
	\includegraphics[width=6in]{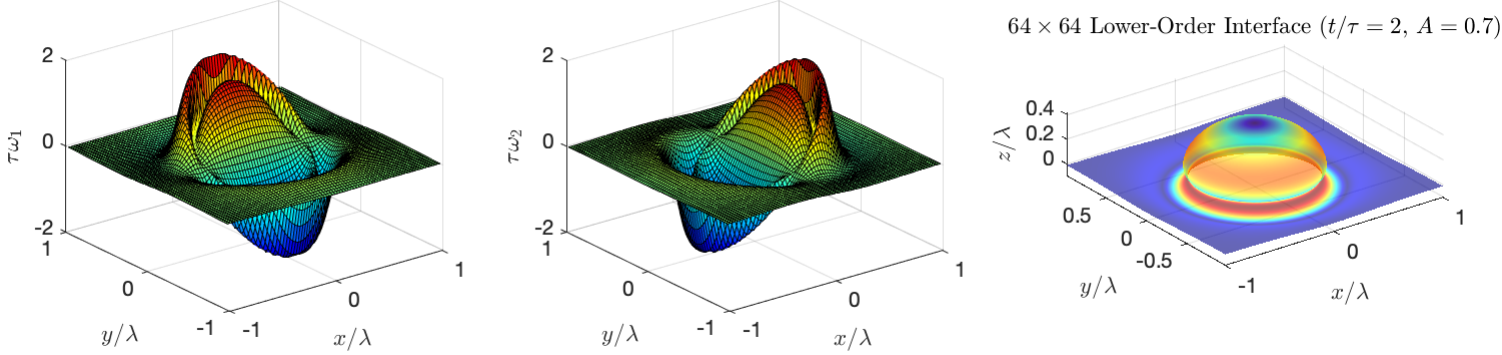}
	\caption{{\footnotesize Nondimensionalized components of vorticity for the lower-order $z$-model, compared with the full interface. Color on the interface indicates magnitude of vorticity (red is higher, blue is lower).}}
	\label{vorticity-lower}
\end{figure}

The lower-order $z$-model fails to roll-up because only one {\it sufficiently strong} ring of vorticity can form during the entire evolution. 
A secondary (weaker) ring of vorticity may form inside the first, but its amplitude is too small to initiate secondary turnover. This weak secondary vorticity produced by the lower-order $z$-model is shown in Figure \ref{vorticity-lower}.

We can now provide a heuristic argument as to why the problematic time-derivative term $2A\partial_t(\bar{\boldsymbol u}\bcdot\p_\alpha\boldsymbol z)$ on the right side
of \eqref{fe2} does not significantly alter the true dynamics, and thus allows for the higher-order $z$-model evolution to accurately simulate solutions of the Euler equations.
Integrating \eqref{fe2}, we have that
\begin{align} 
\mu_\alpha(s, t)  & =  \mathring\mu_\alpha(s) 
+ A \int_0^t \partial_\alpha\big(|\bar{\boldsymbol u}|^2-\tfrac{1}{4}h^{\beta\gamma}\mu_\beta  \mu_\gamma-2gz^3\big) \mathrm{d}t'  \notag \\
&  \qquad \qquad \qquad + 2A \big( \bar{\boldsymbol u}^\epsilon(s,0) \bcdot \p_\alpha \mathring z (s) - \bar{\boldsymbol u}(s,t)\bcdot\p_\alpha\boldsymbol z(s,t) \big) \,.  \label{h-arguement1}
\end{align} 
 From  \eqref{series1} and
the definition of $\bar{\boldsymbol u}$ given in \eqref{u-bar}, we see that for $\alpha =1,2$, 
\begin{align*} 
\bar{\boldsymbol u} \bcdot \p_\alpha\boldsymbol z
&= \iint_{\mathbb{R}^2  }
\frac{(s^\nu-s'^\nu)(s^\beta-s'^\beta)}{|s-s'|^3}\Big(\mu_\nu(s')
\sqrt h\partial_\beta\boldsymbol n
-\tfrac{1}{2}(\mu_2(s')\p_1\boldsymbol z\\
&\hspace{5cm}-\mu_1(s')\p_1\boldsymbol z)\times\boldsymbol n(\p_{\nu\beta}z\bcdot n)\Big) \bcdot  \p_\alpha\boldsymbol z  \mathrm{d}s' \\
&
=-\iint_{\mathbb{R}^2  }\frac{(s^\nu-s'^\nu)(s^\beta-s'^\beta)}{|s-s'|^3}\Big(\mu_\nu(s')
h^{\delta \nu}\p_\delta z\\
&\hspace{4.5cm}+\tfrac{1}{2}(\mu_2(s')\p_1\boldsymbol z-\mu_1(s')\p_1\boldsymbol z)\times n\Big) \bcdot  \p_\alpha\boldsymbol z \ (\p_{\nu\beta}\boldsymbol z\bcdot\boldsymbol n)\mathrm{d}s',
\end{align*} 
where we have used the identity
$$
\p_\beta\boldsymbol n = - h^{\delta \nu} ( \p_{\nu \beta}z \bcdot\boldsymbol n) \p_\delta\boldsymbol z \,,
$$
and the fact that
$$
|s-s'|^3= \big(h_{\nu\beta}(s)(s^\nu-s'^\nu)(s^\beta-s'^\beta)\big)^\frac{3}{2} \,.
$$
Since the inner-product of the first-order term in \eqref{series1} with $\p_\alpha\boldsymbol z$ vanishes, 
$$
\frac{(s^\nu-{s'}^\nu)}{ |s-s'|^3} \mu_\nu(s')\sqrt{h(s)}\boldsymbol n(s)\bcdot \p_\alpha\boldsymbol z =0\,,
$$
the second-order correction is thus given by
\begin{align} 
\frac{(s^\nu-s'^\nu)}{|s-s'|^3} (s^\beta-s'^\beta)\Big(\mu_\nu(s')
h^{\delta \nu}\p_\delta\boldsymbol z
+\tfrac{1}{2}(\mu_2(s')\p_1\boldsymbol z-\mu_1(s')\p_1\boldsymbol z)\times\boldsymbol n\Big) \bcdot  \p_\alpha\boldsymbol z \ (\p_{\nu\beta}\boldsymbol z\bcdot\boldsymbol n) \,.   \label{h-arguement2}
\end{align} 
Now, either the curvature of $\Gamma(t)$ is small in which case $\p_{\nu\beta}\boldsymbol z\bcdot\boldsymbol n$ is small and hence $\bar{\boldsymbol u} \bcdot \p_\alpha\boldsymbol z$ is small, or the
curvature is large, in which case  the vorticity measure forms highly concentrated peaks (as seen in Figure \ref{vorticity-components}), in which the support
of the peaks, which is a good approximation for $|s^\beta-s'^\beta|$, is very small.   Hence, we see from \eqref{h-arguement2} that either small curvature
or strongly concentrated vorticity peaks produces small tangential velocity components.

\section{Numerical Simulation and Validation against Experiments}
\label{sec::validation}

In this section we describe the numerical method for solving the $z$-models, simulate a variety of classical test problems in the Rayleigh-Taylor instability, and validate our higher-order $z$-model to against experimental data of single-mode and random multi-mode RTI.

\subsection{Numerical Approximation of the $z$-Model}
\label{sec::numerics}

The boundary integral formulation of the Euler equations, relying on a parametrization of the material interface, has several attractive features for simulating RTI. 
It is naturally adaptive, concentrating grid points near regions of physical interest, it avoids the numerical diffusion of grid-based methods, and it involves fewer independent variables. 
However, these benefits come at a significant computational cost, due to the nonlinear and nonlocal form of the time derivatives in this formulation. 
The use of standard time-stepping algorithms for evolution equations in which the time-derivatives are linear and isolated cannot be used.  
Rather, simulating such equations requires either the solution of a nonlinear Fredholm integral equation of the second kind for the time-derivatives, or the use of an iterative predictor-corrector method to evolve forward in time. 
Additionally, because the nonlocal expression requires $O(N^2)$ operation to evaluate (where $N$ is the number of interface points), such predictor-corrector methods extremely expensive. 
The higher-order $z$-model has nearly all the aforementioned benefits of the full Euler equations in boundary integral form, and can be solved numerically with straightforward finite differencing in space and Runge-Kutta integration in time, with only a small cost in accuracy (see Section \ref{sec::vorticity}). 

Geometric complexity is an essential feature of RTI, especially in three space dimensions, so accurately simulating such behavior requires maintaining a sharp discontinuity in density across the interface, while at the same time faithfully representing the interface geometry at a subgrid scale. The rectangular cells of Eulerian differencing schemes are ill-suited to the highly curved geometry of RTI problems, and often introduce spurious KHI spirals which obscure the form of the primary instability. Indeed, a look at the bewildering variety of behaviors observed in the excellent comparative study of \citet{LiWe2003} demonstrates that high-order methods such as WENO and PPM, set on structured Cartesian meshes, can be susceptible to spurious small-scale KHI. Nonetheless, for problems which must take into account many physical effects, such as compressibility, thermodynamics, combustion, or electromagnetism, methods such as these provide the best available numerical methods. See \citet{Zh2021} \S5-6 for a recent overview of numerical approaches to RTI. Methods such as marker-and-cell \citep{Da1967} and volume-of-fluid \citep{HiNi1981} use either marker particles or a marker function in addition to an Eulerian grid to track the interface. Similarly, front-tracking methods, which use a triangular mesh that advects with the fluid velocity, have shown considerable success \citep{TrUn1991,Gl1998}.

We now describe our numerical method for the $z$-models. All computations are set on a uniform spatial grid $\{(i\delta,j\delta):i,j=-n,\dots,n-1,n\}$, and the dynamical variables are stored as matrices of values on this grid, e.g.,
\begin{align*}
	\boldsymbol z_{ij}(t)=\boldsymbol z(i\delta,j\delta,t),&&i,j=-n,\dots,n-1,n.
\end{align*}
Spatial derivatives are computed using fourth-order centered-difference operators $(\boldsymbol D_1,\boldsymbol D_2)$. These are used to compute the discretized form of the normal vector $\boldsymbol N=(\boldsymbol D_1\boldsymbol z)\times(\boldsymbol D_2\boldsymbol z)$ and the metric determinant $|h|=|\boldsymbol N|^2$.

For the lower-order $z$-model,  we compute $\widetilde{\boldsymbol u}$ via the fast Fourier transform:
$${\widetilde{\boldsymbol u}}_{jk}=\mathscr F^{-1}\left(\frac{{\mathrm{i}}l({\mathscr F}\mu_1)_{lm}+{\mathrm{i}}m({\mathscr F}\mu_2)_{lm}}{(l^2+m^2)^{\sfrac{1}{2}}}\right)_{jk}\frac{{\boldsymbol N}_{jk}}{|h|_{jk}^{3/2}} \,. $$
This still leaves $({\mathscr F}{\widetilde{\boldsymbol u}})_{00}$ undefined, so we set it to zero. To mitigate the large spikes in vorticity discussed in Section \ref{sec::vorticity}, 
we introduce a nonlinear artificial viscosity operator $\nu \bnabla\bcdot\big(c\bnabla/\sup c\big)$  on the right side of  the vorticity equations, where
$$c=(1-\nu\delta^2\Delta)^{-1}|\boldsymbol\omega| \,, $$
where $\delta$ is the grid size and $\nu >0$ denote the artificial viscosity parameter.
The function $c$ is a  smoothed version of the magnitude of vorticity $|\boldsymbol\omega|$, which provides a smooth localization for the addition of nonlinear artificial viscosity,  
 motivated by the nonlinear (space-time smooth) artificial viscosity method introduced in \citet{ReSeSh2013} and utilized in \citet{RaReSh2019a, RaReSh2019b} but 
employing an elliptic solver at each time-step rather than the solution of a  parabolic reaction-diffusion equation. 
This yields the following set of ODEs for the discretized lower-order $z$-model: 
\begin{align*}
	\frac{\mathrm{d}\boldsymbol z_{jk}}{\mathrm{d}t}&=\widetilde{\boldsymbol u}_{jk} \,, \\
	\frac{\mathrm{d}\mu_{jk}}{\mathrm{d}t}&=A\left[\binom{\boldsymbol D_1}{\boldsymbol D_2}\left(|\widetilde{\boldsymbol u}|^2-\tfrac{1}{4} h^{\alpha\beta}\mu_\alpha\mu_\beta-2gz^3\right)\right]_{jk}+\nu\left[\sum_{\alpha=1}^2\boldsymbol D_\alpha\left(\frac{c\boldsymbol D_\alpha\mu}{\sup c}\right)\right]_{jk} \,,
\end{align*}
which are solved using a total variation decreasing (TVD) third-order Runge-Kutta method in time. For a generic system of ODEs $d\boldsymbol x/dt=\boldsymbol f(\boldsymbol x,t)$, the three-step Runge-Kutta method we use to advance the solution from $\boldsymbol x_n=\boldsymbol x(t)$ to $\boldsymbol x_{n+1}=\boldsymbol x(t+\Delta t)$ is given by
\begin{align*}
	\boldsymbol k_1&=\boldsymbol x_n+\Delta t\boldsymbol f(\boldsymbol x_n,t)\\
	\boldsymbol k_2&=\frac{3}{4}\boldsymbol x_n+\frac{1}{4}\boldsymbol k_1+\frac{\Delta t}{4}\boldsymbol f(\boldsymbol k_1,t)\\
	\boldsymbol x_{n+1}&=\frac{1}{3} \boldsymbol x_n+\frac{2}{3}\boldsymbol k_2+\frac{2\Delta t}{3}\boldsymbol f\left(\boldsymbol k_2,t+\tfrac{1}{2}\Delta t\right)
\end{align*}

For  both the medium and higher-order $z$-models, we compute the Birkhoff-Rott velocity $\bar{\boldsymbol u}$ using a Simpson's rule quadrature, which has 
the form
\begin{equation*}
	\bar{\boldsymbol u}^\epsilon_{ij}=\tfrac{1}{4\pi}\sum_k\sum_lM_{kl}\boldsymbol\omega_{kl}\times\frac{\boldsymbol z_{ij}-\boldsymbol z_{kl}}{(\epsilon^2+|\boldsymbol z_{ij}-\boldsymbol z_{kl}|^2)^{3/2}},\qquad\boldsymbol\omega_{kl}=\big(\mu_{2,kl}(\boldsymbol D_1\boldsymbol z)_{kl}-\mu_{1,kl}(\boldsymbol D_2\boldsymbol z)_{kl}\big) \,,
\end{equation*}
for some matrix of weights $\boldsymbol M=(M_{ij})$. The discretized medium-order $z$-model is given by
\begin{align*}
	\frac{\mathrm{d}\boldsymbol z_{jk}}{\mathrm{d}t}&=\bar{\boldsymbol u}^\epsilon_{jk} \,, \\
	\frac{\mathrm{d}\mu_{jk}}{\mathrm{d}t}&=A\left[\binom{\boldsymbol D_1}{\boldsymbol D_2}\left(|\widetilde{\boldsymbol u}|^2-\tfrac{1}{4}h^{\alpha\beta}\mu_\alpha\mu_\beta-2gz^3\right)\right]_{jk}+\nu\left[\sum_{\alpha=1}^2\boldsymbol D_\alpha\left(\frac{c\boldsymbol D_\alpha\mu}{\sup c}\right)\right]_{jk} \,,
\end{align*}
and the discretized higher-order $z$-model is
\begin{align*}
	\frac{\mathrm{d}z_{jk}}{\mathrm{d}t}&=\bar{\boldsymbol u}^\epsilon_{jk} \,, \\
	\frac{\mathrm{d}\mu_{jk}}{\mathrm{d}t}&=A\left[\binom{\boldsymbol D_1}{\boldsymbol D_2}\left(|\bar{\boldsymbol u}^\epsilon|^2-\tfrac{1}{4}h^{\alpha\beta}\mu_\alpha\mu_\beta-2gz^3\right)\right]_{jk}+\nu\left[\sum_{\alpha=1}^2\boldsymbol D_\alpha\left(\frac{c\boldsymbol D_\alpha\mu}{\sup c}\right)\right]_{jk} \,.
\end{align*}
\begin{figure}
	\centering
	\includegraphics[width=4in]{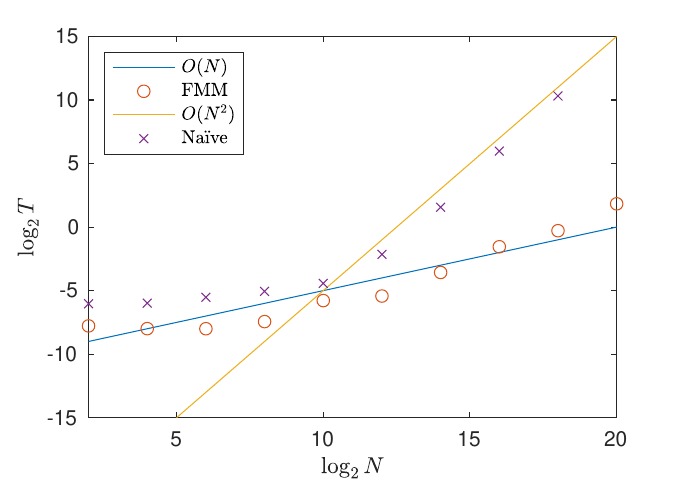}
	\caption{{\footnotesize Measured computation times for the Birkhoff-Rott velocity.}}
	\label{benchmark}
\end{figure}

The  time required for the computation of the Birkhoff-Rott velocity $\bar{\boldsymbol u}$ can be drastically shortened with the use of the fast multipole method. 
Evaluating $\bar{\boldsymbol u}^\epsilon_{jk}$ for all $j$ and $k$ requires $O(N^2)$ operations, where $N=(2n+1)^2$ is the total number of grid points in the simulation. 
In contrast, the fast multipole method evaluates in $O(N)$ operations, given a fixed relative accuracy goal.  As shown in Figure \ref{benchmark}, we have 
verified this performance for interface grid resolutions up to size $N=1024\times1024$ using the open-source package 
\texttt{fmm3d}.\footnote{https://github.com/flatironinstitute/FMM3D}

\subsection{Convergence of the $z$-Model}
\label{sec::convergence}

We note that existence of solutions to the  incompressible Euler equations with general vortex sheet initial data is not known, and for certain types
of vortex sheet data for which weak solutions can be constructed, such solutions are not unique \citep[see, for example][]{Sz2011}.   As such
 different regularizations of the Euler equations will lead to sequences of approximate solutions which, such that if these sequences converge, they
 may not all converge to the same solution.  
Sequences of approximate Euler solutions, obtained by using a regularized Birkhoff-Rott velocity, have been shown to converge to (possibly
non-unique) solutions of the Euler equations in, for example, 
\cite{BeMa1982,LiXi1995,LiXi2001}.   In particular, numerical simulations suggest  that the limiting weak solution depends upon the choice of 
the velocity regularization \citep[see][]{BaPh2006,FiLoLoZh2006,RaSh2020}.

On the other hand, 
as was shown in \citet[Figures 8-10]{RaSh2020}, the 2D $z$-model exhibits convergence of  large-scale averaged quantities such as the spiral
roll-up center, the roll-up radius, the location of the bubble and spike tips, and the bubble and spike amplitudes.  It is likely that such convergence of
large-scale features is independent of the numerical approximation employed, and 
in this section we show that the 3D higher-order $z$-model converges in this large-scale sense,  in the limit as the desingularization and mesh spacing are taken to zero simultaneously.

\begin{figure}
	\centering
	\includegraphics[width=2.5in]{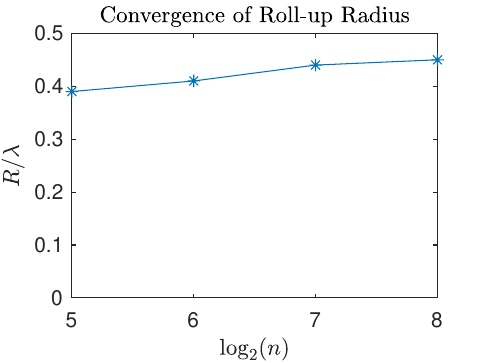}
	\includegraphics[width=2.5in]{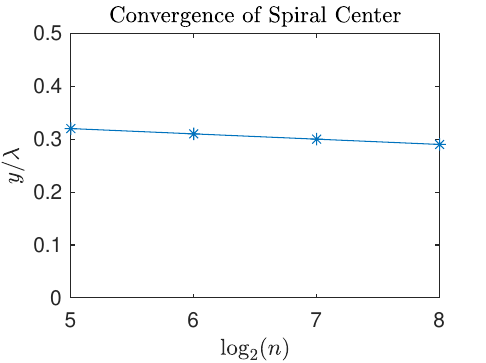}
	\caption{{Convergence of roll-up radius and spiral centers for the 3D higher-order $z$-model, as the mesh is refined.}}
	\label{convergence}
\end{figure}

Figure \ref{convergence} demonstrates this large-scale  convergence for the  3D higher-order $z$-model , using the initial data \ref{gaussian}.  As
can be seen,  there is convergence of the roll-up radius and  the location of the spiral centers on a sequence of $n\times n$ meshes with $n=32,64,128,256$, and with regularization parameter $\epsilon$ proportional to the mesh size.

\subsection{Single-Mode Initial Data}
\label{sec::single-mode-data}

Sinusoidal perturbations of a flat interface were the first to be studied in the context of RTI, and much of the analytical and experimental study of RTI has focused on understanding the evolution of single-mode initial 
data beyond the linear regime \citep[see, for instance, ][]{ChEmWa1960,CoTa1973,BaMeOr1984,Go2002}. 
Of course, in most application scenarios, the interface is perturbed with many different frequencies, but to understand the single-mode problem is the first step towards understanding the more general problem.

\begin{figure}
	\centering
	\includegraphics[width=6in]{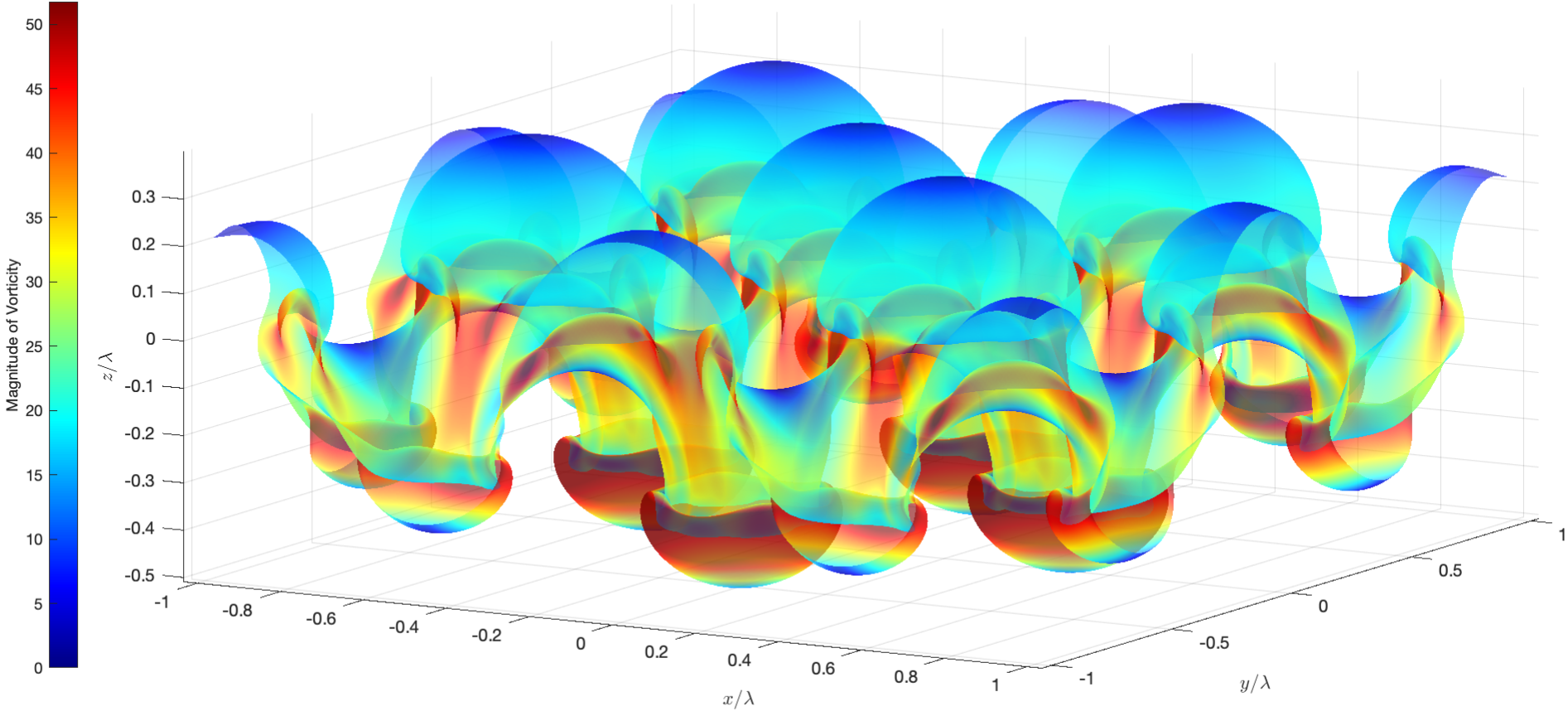}
	\caption{{Single-mode initial data with two wavelengths along each side of the tank, run using the higher-order $z$-model on a $192\times192$ grid, with Atwood number 0.15 and a computation time of 380s. }}
	\label{onemode}
\end{figure}

For our first test, we use our higher-order $z$-model to simulate Wilkinson/Jacobs' isopropyl alcohol/water experiment ($A=0.15$). We initialize our $z$-model with initial conditions of the form
\begin{equation}\label{eqn::singlemode-IC}
	z^1=s^1,\quad z^2=s^2,\quad z^3=a_0\cos(\pi k s^1/\lambda)\cos(\pi k s^2/\lambda),\quad \mu_1=\mu_2=0,
\end{equation}
where $-\lambda\leq s^1,s^2\leq\lambda$ and $k$ is a positive integer.   We may regard $\lambda$ as a characteristic length for this problem, and the corresponding characteristic time scale is given by $\tau=(\lambda/Ag)^{\sfrac{1}{2}}$.  This type of initial data was studied experimentally by Wilkinson and Jacobs,
 who used planar laser-induced fluorescence (PLIF) to visualize a diagonal cross-section of the fluid as it evolved, showing the 
{\it double roll-up} which distinguishes 3D Rayleigh-Taylor instability from 2D \citep{JaWi2007}.
Results for the case $k=2$ are shown in Figure \ref{onemode}.

Starting from a single-mode perturbation of a flat interface, 3D RTI forms a periodic pattern of bubbles and spikes in such a way that \emph{two} vortices form on the side of each bubble or spike. In 2D, single-mode RTI produces a stationary row of counter-rotating vortices, which manifests as a single row of roll-up regions. However, in 3D, single-mode RTI produces two grids of counter-rotating vortex \emph{rings}, which move apart from each other at roughly constant speed. 
Upon taking a cross-section along the diagonal of the tank, we obtain the double-spiral pattern seen here. 
See \citet{ChJa2006} Figures 8 and 16 for a helpful illustration of single vs. double roll-up regions.

\begin{figure} 
	\centering
	\includegraphics[width=6in]{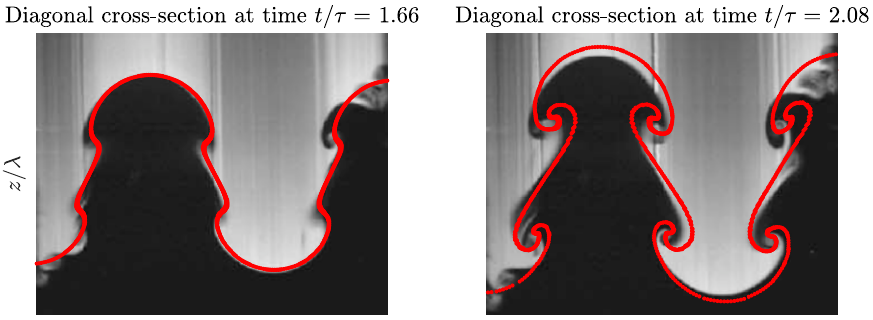}
	\includegraphics[width=6in]{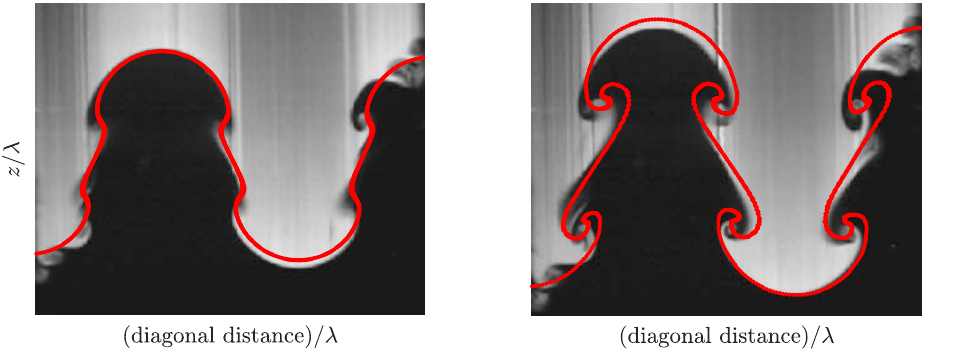}
	\caption{ {\footnotesize \textit{Top:} Higher-order $z$-model. \textit{Bottom:} Medium-order $z$-model. Both were run on $160\times160$ meshes, with Atwood number 0.15 and a runtime of 270s. Experimental photographs from  \cite{JaWi2007}. Copyright 2007, AIP Publishing. Reproduced with permission. }}
	\label{onemode_comparison}
\end{figure}
\begin{figure} 
	\centering
	\includegraphics[width=6in]{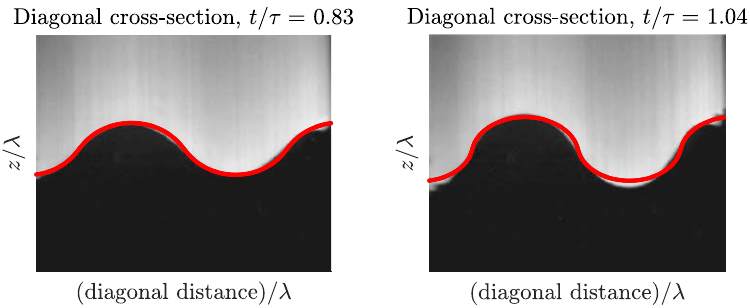}
	\caption{{\footnotesize Lower-order $z$-model, run on a $130\times130$ grid, with Atwood number $0.15$ and a runtime of $9$s. Experimental photographs from \citet{JaWi2007}. Copyright 2007, AIP Publishing. Reproduced with permission.}}
	\label{onemode_lo}
\end{figure}

Using the parameters from their experiments and matching the wavelength of the initial data as best we could, we were able to achieve remarkably good agreement between our higher- and medium-order $z$-models and their experimental data -- see Figure \ref{onemode_comparison}. As shown in Figure \ref{onemode_lo}, the lower-order $z$-model matches the experiments quite well at early times.  At later times,  however, the lower-order
$z$-model cannot capture the full extent of the roll-up, as discussed in Section \ref{sec::vorticity}.

\begin{figure}
	\centering
	\includegraphics[width=4in]{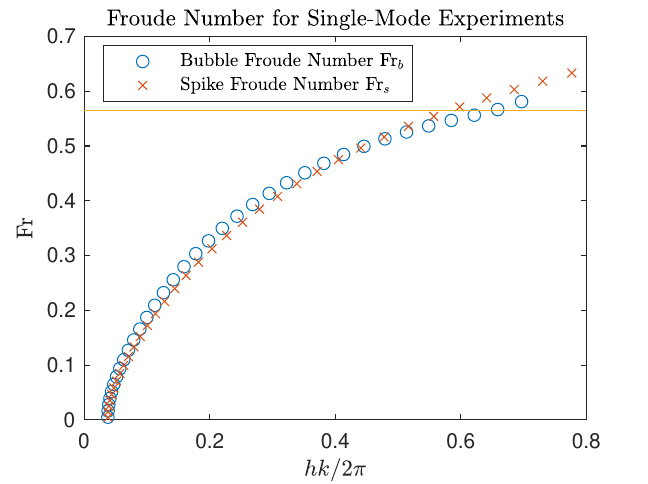}
	\caption{{\footnotesize Plot of bubble and spike Froude numbers against normalized amplitude for our single-mode RTI simulations. Horizontal line shows asymptotic prediction $\text{Fr}\sim\pi^{-\sfrac{1}{2}}$. Compare with Figs 15 and 16 in \citet{JaWi2007}. }}
	\label{froude}
\end{figure}

The amplitude of 3D single-mode Rayleigh-Taylor instability initially follows the exponential growth rate $a\sim\cosh(\sqrt{Agk}t)$ predicted by the linear theory, before slowing to approximately constant velocity. The single-mode model of \citet{Go2002} predicts asymptotic bubble and spike velocities of
$$U_b=\sqrt{\frac{2Ag}{(1+A)k}},\qquad\qquad U_s=\sqrt{\frac{2Ag}{(1-A)k}},$$
respectively, where $k$ is the wavenumber of the initial perturbation. However, an asymptotically constant velocity has been called into question by numerical calculations of \citet{CaDiFrRaYo2006}. Nonetheless, the single-mode experiments of \citet{JaWi2007} show a period of linear growth in amplitude near Goncharov's predicted value, before accelerating once again. In Figure \ref{froude} we have plotted the Froude number (non-dimensional bubble velocity) for the bubbles and spikes in our single-mode experiments, 
$$\text{Fr}_b=\frac{\dot a_b}{\sqrt\pi U_b},\qquad\qquad\text{Fr}_s=\frac{\dot a_s}{\sqrt\pi U_s}$$
against the normalized bubble and spike amplitudes $a_b$ and $a_s$. Although our simulations do not run for as long as Wilkinson and Jacobs' experiments, the Froude numbers are in agreement with the experimental results for the simulated range of amplitudes, showing an increase to slightly above the predicted asymptotic value $\text{Fr}\sim\pi^{-\sfrac{1}{2}}$ \citep[compare our Figure \ref{froude} with Figures 15-17 in][]{JaWi2007}.

\subsection{The 3D Rocket Rig Experiment}
\label{sec::rocketrig}

The growth rate of the mixing layer between two fluids subjected to random multi-mode perturbations has been the subject of significant interest and study, and it forms a useful benchmark for our $z$-models. Starting with a small perturbation of amplitude $h_0$ of a flat interface, theory predicts that the amplitude $h(t)$ of the resulting mixing layer has the form
$$
\frac{h-h_0}{\lambda}=\alpha\left(\frac{t}{\tau}\right)^2 \,,
$$
where $\alpha$ is a universal growth rate constant. The growth rate of the interface was tested experimentally by \citet{Re1984} and \citet{Yo1989}, and numerically by \citet{Yo1984,Yo1989}.  
Their experimental setup consists of a $150\times150\text{mm}^2$ tank on a vertical sled, accelerated downward using rocket motors, which gives this experimental setup the nickname ``rocket rig''.   
Read's results suggest that $\alpha$ is between $0.06$ and $0.07$, although individual experiments have given values as low as $0.058$ and as high as $0.077$.  
To test the higher-order $z$-model in this scenario, we simulate Read's experiment using sodium iodide (NaI) solution $(\rho^+=1.9\text{ g/cm}^3)$ and pentane $(\rho^-=0.63\text{ g/cm}^3)$ , which results in an Atwood number $A\approx 0.5$. 
More recently, \citet{ClRi2004} and \citet{CaCoMi2004} independently discovered a new governing equation for the mixing layer amplitude, 
$$\dot h^2=C_0Agh.$$
At large times, the solution to this ODE grows like $h\sim\frac{1}{4}C_0t^2$, so that $C_0=4\alpha$. However, it was found by \citet{CaCo2006} that the ratio $\dot h^2/4Agh$ approaches $\alpha$ only at very long times (around $t=6$-$7\tau$, more than double the runtime of the experiments of Read and Youngs). The numerical experiments of \citet{CaCo2006} ran until $t=30\tau$, or more than eleven times the runtime of Read's experiments, and found that $\alpha$ stabilized at a value around $0.025$. Our numerical experiments were run only for the length of Read's experiments, but the timeseries plot of $\dot h^2/4Agh$ from our experiments matched that of Cabot \& Cook for the length of our experiments.

\begin{figure}
	\centering
	\includegraphics[width=3in]{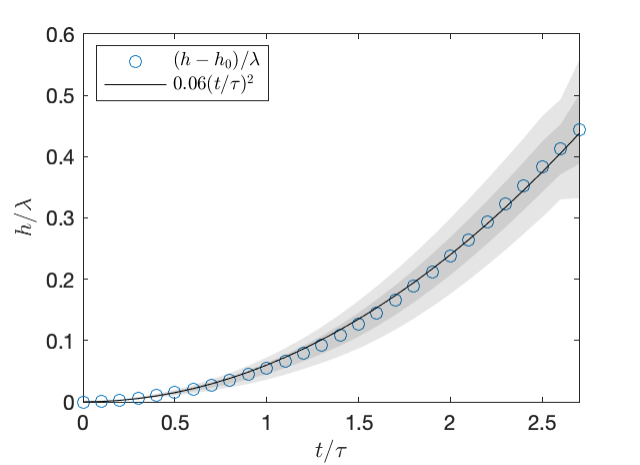}\hspace{-5mm}
	\includegraphics[width=3in]{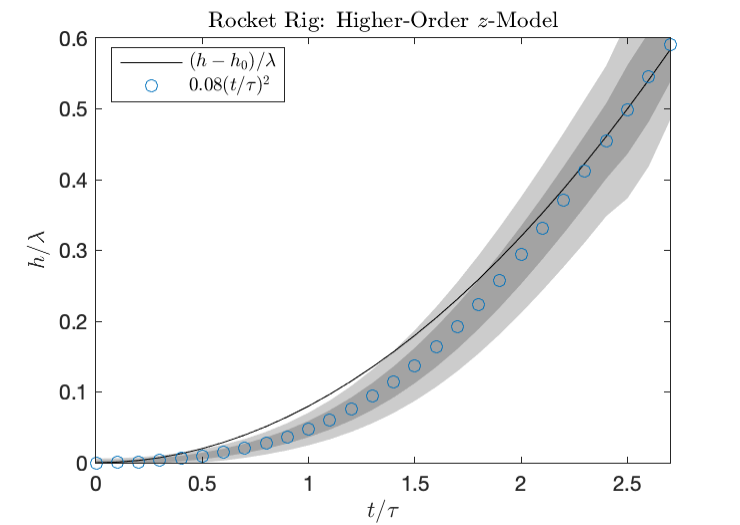}
	\caption{{\footnotesize Comparison of higher-order $z$-model with theoretical growth rate, Case A (\textit{Left}) and Case B (\textit{Right}). The markers show the amplitude growth of the interface, averaged over 100 randomly initialized runs, and the gray bands show one and two standard deviations. }}
	\label{growth}
\end{figure}
\begin{figure} \label{rr_lo}
	\centering
	\includegraphics[width=3in]{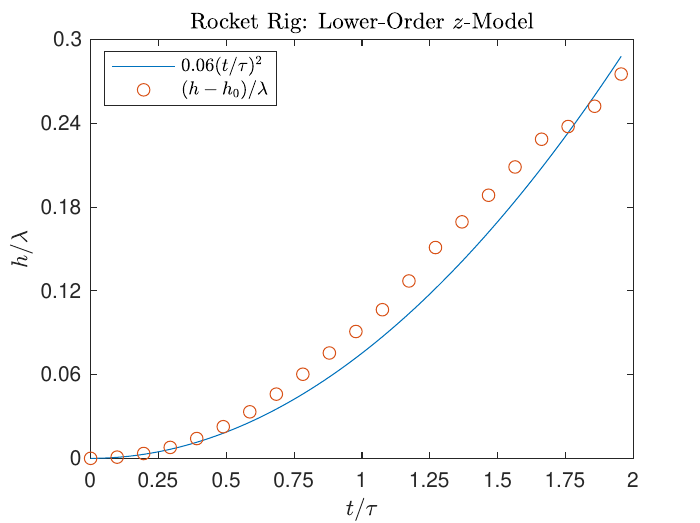}
	\caption{{\footnotesize  Comparison of lower-order $z$-model with theoretical growth rate, initialized in Case B. }}
	\label{growth_lo}
\end{figure}
In this article we have simulated the rocket rig using parameter values from Read's NaI/Pentane experiment. 
Two main strategies are possible to initialize the rocket rig for numerical simulations, representing two distinct regimes of instability. The first strategy, Case A, is to initialize the instability with a random combination of short-wavelength perturbations, and allow larger scale disturbances to develop from the nonlinear interaction of the short-wavelength perturbations. The second strategy, Case B, is for long-wavelength perturbations to be present in the initial condition, causing the mixing layer to grow more rapidly at early times. 
See \citet{Yo2013} for a more detailed discussion and comparison of the different initialization strategies for the rocket rig. In Case A, we initialize our model with a random perturbation of a flat interface, of the form
$$z_1=s_1,\qquad z_2=s_2,\qquad z_3=A_0\text{Re}\sum_{j_1=-n}^n\sum_{j_2=-n}^na_{j_1,j_2}e^{\mathrm{i}(j_1s_1+j_2s_2)}$$
where the complex coefficients $a_{\boldsymbol j}=a_{j_1,j_2}$ are drawn from a standard normal distribution when $|\boldsymbol j|>n/2$ and zero otherwise, and $A_0$ is chosen so that $(\iint|z_3(s,0)|^2\mathrm{d}s)^{\sfrac{1}{2}}=0.05$. In Case B, we instead draw the coefficients $a_{\boldsymbol j}$ from a normal distribution with variance $\sigma^2(\boldsymbol j)\propto|\boldsymbol j|^{-3}$.

\begin{figure} 
	\centering
	\includegraphics[width=3in]{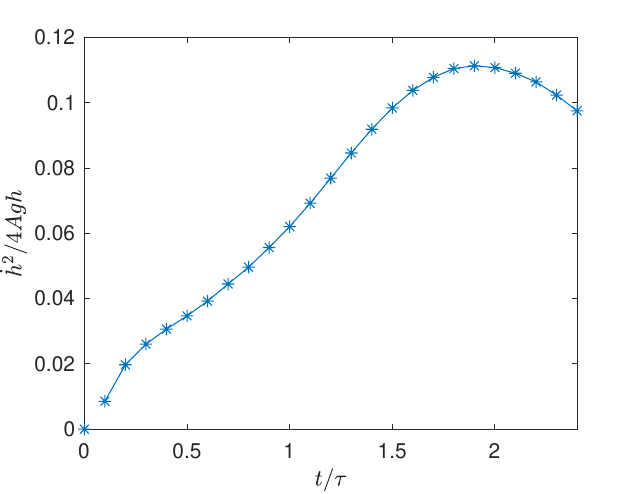}
	\caption{{\footnotesize   Plot of the time-dependent growth rate $\dot h^2/4Agh$, averaged over 100 rocket rig runs.}}
	\label{ristorcelli}
\end{figure}
Figure \ref{growth} shows results from 100 simulations of the higher-order $z$-model, using high-frequency initial perturbations (Case A, left panel) or the `enhanced mixing' nonuniform variance (Case B, right panel). Both the experiment and the simulation were run for 67.4ms, or $2.7\tau$.  
The Case A runs show excellent agreement with the value $\alpha=0.06$ suggested by Read's experiments. The computation time was 100s/run, using $100\times100$ meshes. The lower-order model does reasonably  well also, but the rate of interface growth slows after about two characteristic times, because the lower order $z$-model is not able to fully capture interface roll-up. This is shown in Figure \ref{growth_lo}. All models achieve growth rates around the range found in Read's experiments $(0.058\leq\alpha\leq0.077)$.
We note that although our simulations do not run long enough to see the stabilization of growth rate $\dot h^2/4Agh\sim\alpha$, they qualitatively match the plot of $\dot h^2/4Agh$ shown in Cabot \& Cook up to the runtime of our simulation, showing an increase up to a peak of about $0.1$ followed by steady decrease \citep[compare Figure \ref{ristorcelli} to][Figure 4]{CaCo2006}.
To our knowledge, these are the first interface models which can accurately reproduce the growth rate of disordered RT mixing layers.

\subsection{Fluid mixing in the Rocket Rig experiment}
\label{sec::mixing}
\begin{figure}
	\centering
	\includegraphics[width=6in]{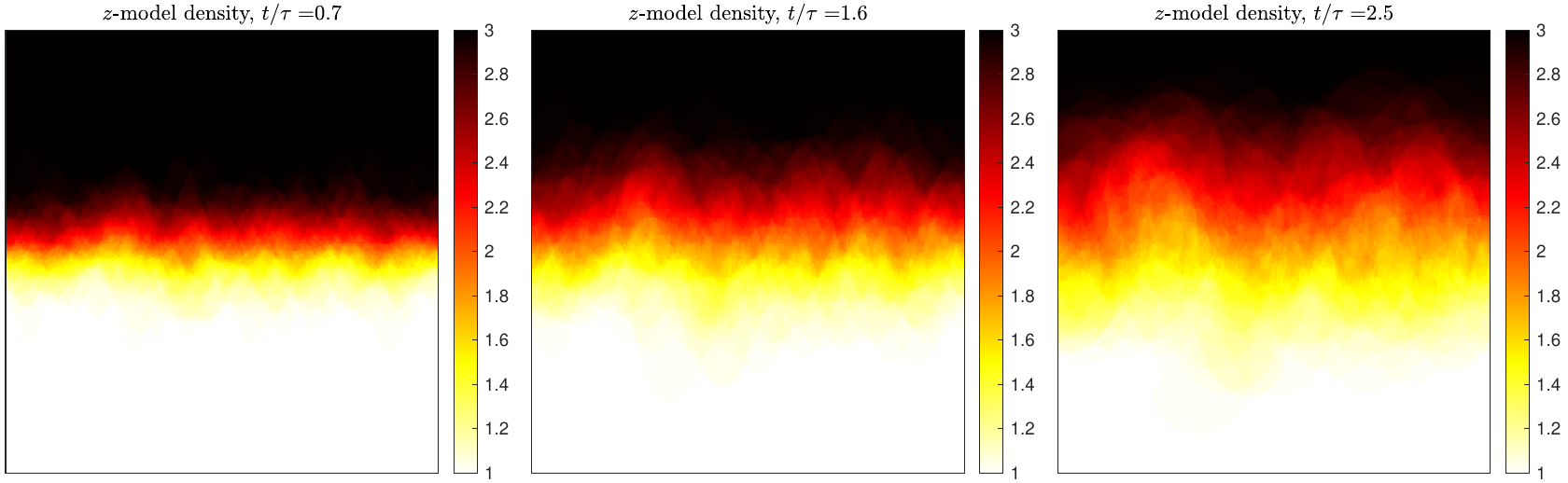}
	\caption{{\footnotesize  Ensemble averaged density for the 100 runs of the rocket rig analyzed above, showing a cross-section at $y=0$.  }}
	\label{rocketrig_mixing}
\end{figure}

In addition to measuring the growth rate of the mixing layer, the speed of our high-order $z$-model simulations makes it feasible to compute ensemble averages of  density for randomized initial data as shown in Figure \ref{rocketrig_mixing} for the case of an ensemble of 100 runs.   When computing ensembles using the invariant (Gibbs) measure associated to the Euler equations, it is known   \citep{AlCr1990} that ensemble averaging of runs with data using (Gaussian) random 
frequency fluctuations of the interface produce viscous effects (whose size is inversely proportional to the frequency of perturbation).    Such viscous effects can be computed
from a uniform ensemble\footnote{For the purpose of numerical simulations, it is much easier to compute the uniform ensemble than the ensemble averaging using the invariant
measure, but the viscous effects are extremely similar.} of runs, and our results for such ensembles (see Figure \ref{rocketrig_mixing})  is in good agreement with traditional simulations of RTI using
the Navier-Stokes equations.

The horizontally-averaged density as a function of the vertical coordinate and time, which we will denote $\bar\rho(z,t)$, is a useful measurement of the large-scale phenomenology of Rayleigh-Taylor turbulence. 
While it is expensive to compute this function using direct numerical simulation (DNS), \citet{BoLiMu2010} showed that the solution to a simple parabolic model equation for $\bar\rho$ using the so-called Prandtl closure model closely matches results from DNS.  
Our objective is to demonstrate that the ensemble-averaged density field computed from 100 $z$-model runs reproduces results obtained from the Prandtl closure model, and thus also reproduces results from DNS.  
In particular, we shall demonstrate that the $z$-model, even
at low resolution, captures the statistical qualities of turbulent mixing quite accurately and reproduces the remarkable self-similarity of the horizontally-averaged density.

Averaging the density transport equation $\partial_t\rho+\partial_x(\rho u)+\partial_y(\rho v)+\partial_z(\rho w)=0$ over the horizontal ($x$ and $y$) coordinates results in the equation $\partial_t\bar\rho+\partial_z\overline{\rho w}=0$, where $\bar\rho$ is the horizontally averaged density field given by
$$\bar\rho(z,t)=\fint_{\mathbb{R}^2}\rho(x,y,z,t)dxdy,$$
and $\overline{\rho w}$ is the horizontally averaged vertical mass flux. To close the equation for $\bar\rho$, the Prandtl closure model  assumes
that  $\overline{\rho w}=\kappa\partial_z\bar\rho$, where $\kappa$ is known as the ``eddy diffusivity''. This eddy diffusivity has units of length$^2$/time, so it
is natural to take $\kappa\propto h\dot h$, where $h$ is a characteristic length and $\dot h$ is a characteristic velocity for the turbulent flow. 
In our case, $h\propto Agt^2$ is the characteristic mixing layer height in the rocket rig, so that
$$\kappa\propto(Ag)^2t^3 \,.$$
Letting $\beta^2$ denote the dimensionless constant of proportionality, the following evolution equation for $\bar\rho$ is obtained \citep{BoLiMu2010}:
\begin{equation} 
\partial_t\bar\rho=(\beta Ag)^2t^3\partial_z^2\bar\rho \,. \label{prandtl-closure}
\end{equation} 
This equation has the similarity  variable $\xi=z/(Agt^2)$, which results in a self-similar solution of the form $\bar \rho(z,t)=P(\xi)$ for the density 
field in the rocket rig experiment given by
\begin{equation}
	P(\xi)=\rho^-+\frac{\rho^+-\rho^-}{2}\left(1+\text{erf}\left(\frac{\xi}{\beta}\right)\right) \,.
	\label{selfsimilarsolution}
\end{equation}
Note that as $t\to 0^+$, $P(\xi)$ approaches the step-function initial density 
$$\bar\rho(z,0)=\begin{cases}
	 \rho^+,&z>0,\\
	 \rho^-,&z\leq 0.
\end{cases}$$
As noted above,   \citet{BoLiMu2010} have shown that the self-similar solution $P(\xi)$ to \eqref{prandtl-closure} is an accurate approximation to the density field $\bar\rho(z,t)$ computed from fully resolved DNS.

\begin{figure}
	\centering
	\includegraphics[width=6in]{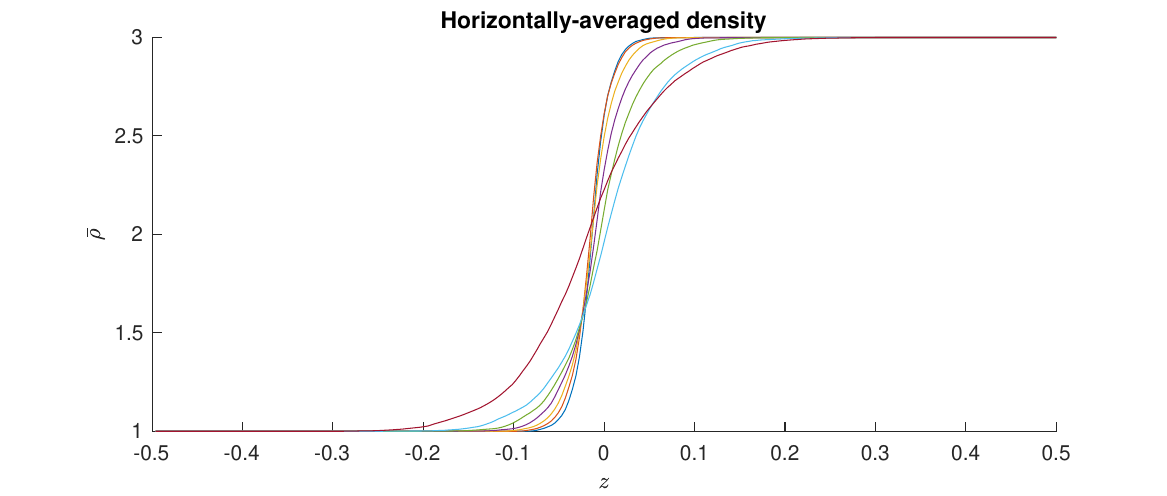}
	\caption{{\footnotesize Horizontally averaged density field $\bar\rho(z,t)$ for the rocket rig experiment, computed from the ensemble-averaged $z$-model density, plotted at times between $t=1.3\tau$ and $t=2.5\tau$, with time intervals $\Delta t=0.2\tau$.}}
	\label{rr_density}
\end{figure}

\begin{figure}
	\centering
	\includegraphics[width=6in]{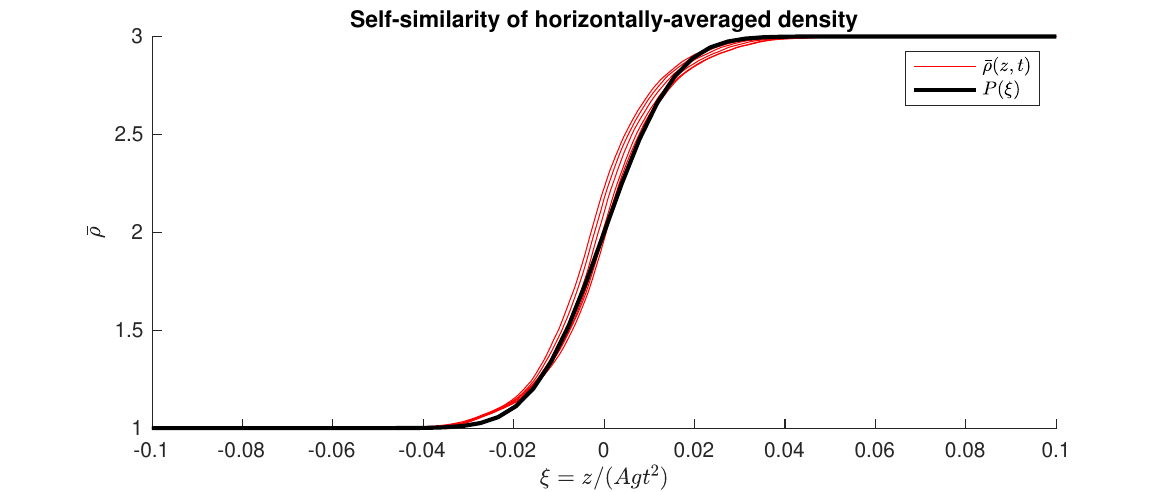}
	\caption{{\footnotesize The same density profiles as in Figure \ref{rr_density}, plotted as a function of the similarity variable $\xi=z/(Agt^2)$ and compared against the theoretical solution $P(\xi)$. 
	}}
	\label{rr_selfsimilar}
\end{figure}

In order to determine if the $z$-model can capture the self-similar solution of the Prandtl closure model, we shall use the ensemble-averaged density
field $\rho$ computed from 100 $z$-model simulations, and then calculate the horizontally-averaged density field $\bar\rho(z,t)$.
We note that the mass associated to the horizontally averaged density field, given by $\int_{-0.5 \lambda}^{0.5\lambda} \bar \rho(z,t) dz$, 
is approximately conserved; in particular, on the full time-interval $0 \le t \le 2.5\tau$ of the numerical experiment, we have verified that mass
deviates from its initial value by less than  $1\%$.
A series of time snapshots $\bar\rho(\cdot,t)$ are plotted in Figure \ref{rr_density} at various times between  $t=1.3\tau$ and $t=2.5\tau$, at intervals $\Delta t=0.2\tau$.
To compare against the self-similar solution $P(\xi)$ in  $\ref{selfsimilarsolution}$, we have plotted the curves 
$\bar\rho(z,t)=\bar\rho(Agt^2\xi,t)$ (for various values of $t$)  as a function of $\xi$ in Figure \ref{rr_selfsimilar}.
Here $\beta$ is chosen to be the ``best-fit'' value for the data, which we found to be approximately  $\beta=0.017$. 
As can be seen in Figure \ref{rr_selfsimilar}, the $z$-model horizontally-averaged density profiles  $\bar\rho(z,t)$ enjoy the self-similar profile
of the solution $P(\xi)$ produced by the  Prandtl closure model, and are thereby compare well with fully resolved DNS.

\begin{figure}
	\centering
	\includegraphics[width=3in]{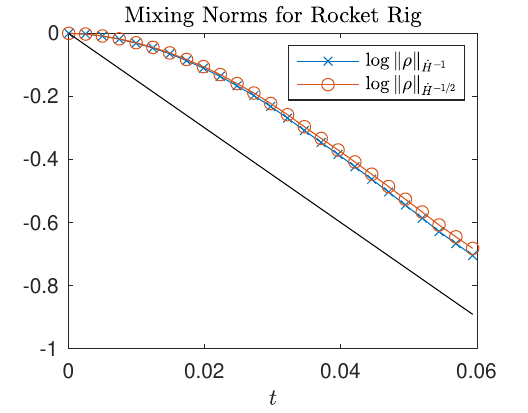}
	\caption{{\footnotesize  Mixing norms of the ensemble-averaged density shown in Figure \ref{rocketrig_mixing}, plotted on a log scale. }}
	\label{mixingnorms}
\end{figure}

We next demonstrate that the ensemble averaging of the $z$-model density field reproduces optimal mixing rates.
As explained in \cite{DoLiTh2011} and \cite{Th2012}, the mixing of a passive scalar (in our case the density $\rho$), advected by a velocity field 
$\boldsymbol u$, can be measured quantitatively using the homogeneous Sobolev norms
$$\|\rho\|_{\dot H^s}=\left(\int_{\mathbb{R}^3}|\boldsymbol k|^{2s}|\hat\rho(\boldsymbol k)|^2\mathrm{d}\boldsymbol k\right)^{\sfrac{1}{2}}$$
with $s<0$. Of particular interest are the so-called {\it mixing norms} corresponding to the exponents $s=-1$ and $s=-\sfrac{1}{2}$.  A fluid  becomes molecularly mixed when the varying density field $\rho(x,y,z,t)$ converges to its spatial average $\langle\rho\rangle=\fint_{\mathbb{R}^3}\rho(x,y,z,t)dxdydz$. 
\citet{DoLiTh2011} showed that $ \rho(\cdot,t)$ can converge at most exponentially fast to its average in one of these mixing norms, and showed that optimal mixing is achieved if the density is advected by one of a few simple, explicit ``stirring'' velocity fields $\boldsymbol u(\boldsymbol x,t)$. However, it is unknown whether optimal mixing is achieved for more general velocity fields, such as those satisfying the Euler or Navier-Stokes equations. That is, optimal mixing occurs if
$$\|\rho( \cdot , t) - \langle\rho\rangle\|_{\dot H^s}\sim e^{-rt}\,, \ \ \ s=-1 \text{ or } s=-\tfrac{1}{2} $$
for some constant $r>0$ as time  $t\to\infty$  \citep[see][]{Th2012}. 
By computing the $\dot H^{-\frac{1}{2}}$ and $\dot H^{-1}$ norms of the ensemble-averaged rocket rig density in a small strip around the plane $z=0$, we find that the Sobolev norms decay exponentially (Figure \ref{mixingnorms}).   
In other words, the ensemble average of $z$-model simulations with randomly perturbed initial data is consistent with optimal mixing of the two fluids. 
While a single $z$-model  simulation  approximates the motion a fluid interface separating two inviscid fluids, the ensemble average of many 
simulations  successfully models viscous mixing at the length scales of the random fluctuations of the data. 
A more in-depth study of mixing rates in Rayleigh-Taylor problems, and in particular whether they achieve optimal mixing, is an objective of our future studies.

\subsection{Fluid Mixing in Complex Stratifications}\label{sec::stratified}

In multiphase fluid problem, in which the interface exhibits significant roll-up and small-scale structure formation, the interface can ``squeeze'' a
fluid into a very thin configuration, as the distance between portions of the fluid interface becomes smaller and smaller.   The
resulting  small distance between portions of the interface creates (super-exponential) growth of the density gradient and ensures that traditional
numerical schemes, based upon a multi-dimensional discretization of the fluid domains, can become  prohibitively expensive due to severe small-scale
resolution requirements.
Such small-scale formation occurs in many situations when significant Kelvin-Helmholtz instability is allowed to develop, or when the fluid becomes turbulent.   Our interface $z$-model is designed specifically for this small-scale scenario, in which  the use of conventional numerics would  not
be feasible. 

\begin{figure}
	\centering
	\includegraphics[width=6in]{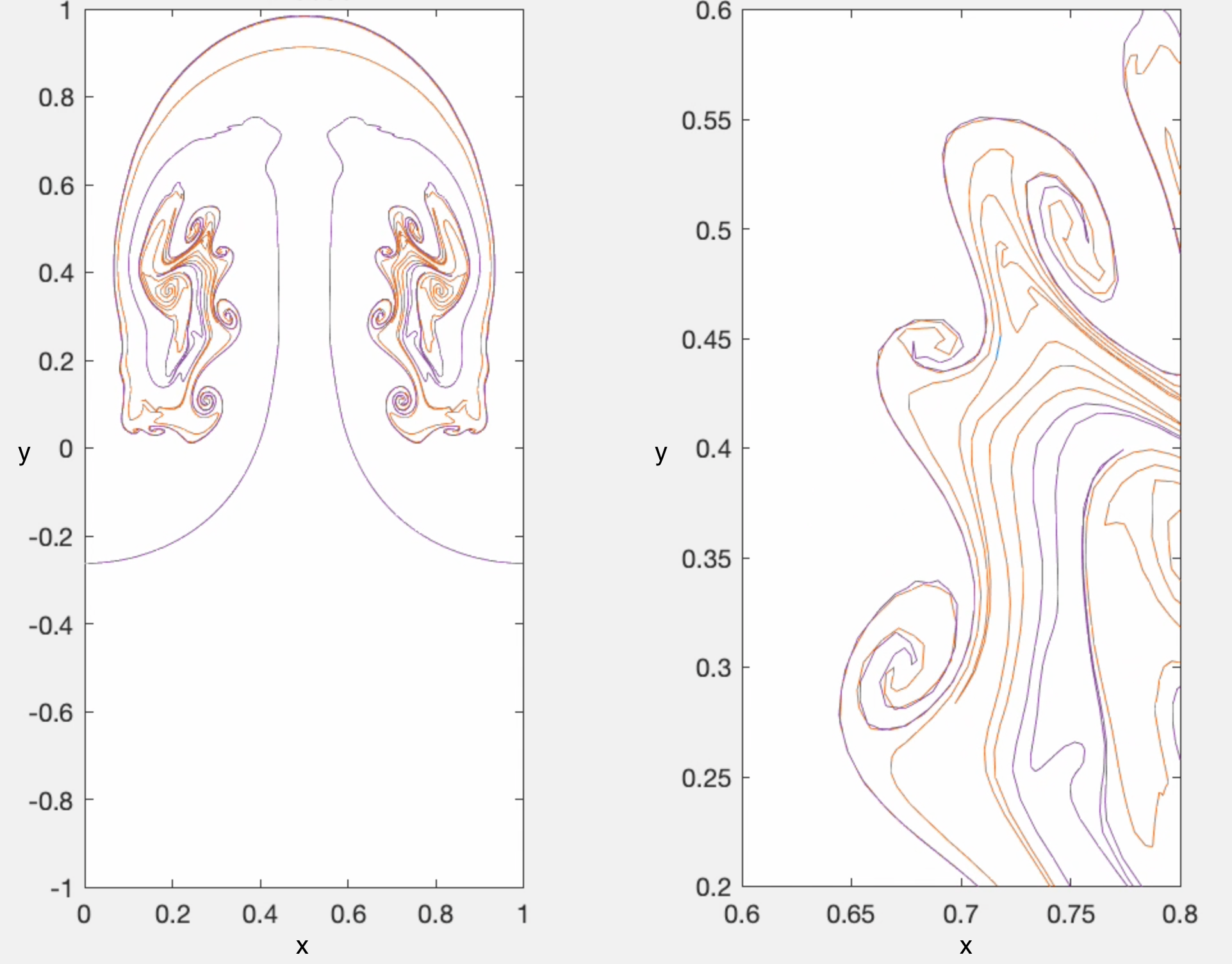}
	\caption{{\footnotesize Simulation of an unstably stratified flow with densities $\rho_1=1.8$, $\rho_2=3$, $\rho_3=5$, with a sinusoidal initial perturbation of the lower interface. (\textit{Left}) The full domain. (\textit{Right}) Zoomed-in view of the Kelvin-Helmholtz instability. }}
	\label{stratified}
\end{figure}

Unstable stratified flows with more than two fluid layers provide an interesting example of such small-scale structure formation. We shall
consider the problem of two fluid interfaces separating three fluids, and our initial data is chosen as
$$\boldsymbol u_0(x,y,z)=\boldsymbol 0,\qquad\qquad\rho_0(x,y,z)=\begin{cases}
	\rho_3,&z>h_2(x,y)\\
	\rho_2,&h_1(x,y)\leq z\leq h_2(x,y)\\
	\rho_1,&z<h_1(x,y)
\end{cases}$$
where $\rho_1<\rho_2<\rho_3$ denote the initial densities of the three fluids, and  $h_1$ and $h_2$ are given functions which represent the
initial position of the two fluid interfaces.
Our $z$-model is easily generalized to simulate such problems by introducing two interfaces $\Gamma_i(t)$, parametrized by $\boldsymbol z_i(s,t)$ and carrying vortex sheet densities $\boldsymbol\mu_i(s,t)$, $i=1,2$. Following our derivation of 
 the standard $z$-model, we obtain that for $i=1,2$, 
\begin{align*}
	\partial_t\boldsymbol z_i(s,t)&=\bar{\boldsymbol u}_i(s,t):=\frac{1}{4\pi}\sum_{j=1}^2\iint_{\mathbb{R}^2}\frac{\mu_{j2}(s',t)\partial_1z_j(s',t)-\mu_{j1}(s',t)\partial_2z_j(s',t)}{|z_i(s,t)-z_j(s',t)|^3}ds' \,, \\
	\partial_t\mu_i(s,t)&=A\nabla_s\left(|\bar{\boldsymbol u}_i(s,t)|^2 -\frac{1}{4}g_i^{jk}(s,t)\mu_{ij}(s,t)\mu_{ik}(s,t)-2gz_i^3(s,t)\right) \,.
\end{align*}
When the two fluid interfaces roll-up inside one another and the separation distance becomes exceedingly small, the
density gradient grows exceedingly large, and traditional numerical schemes based on multi-dimensional grids require
enormous resolution to accurately capture the small-scale features of the interface and maintain stability.  On the other hand,  a small 
interface separation distance does not impair the numerical stability of our interface model, and a two-interface interaction is used to
demonstrate stability under small-scale formation.  We consider 
a two-dimensional unstably stratified three-fluid flow with densities $\rho_1=1.8$, $\rho_2=3$, $\rho_3=5$, and with the two interfaces given a sinusoidal perturbation of wavelength $\lambda$ and amplitude $0.01\lambda$, separated by a distance $0.05\lambda$.
As shown in Figure \ref{stratified}, the unstable stratification accelerates the development of Rayleigh-Taylor instability, resulting in strong Kelvin-Helmholtz roll-up with the distance between the two interfaces becoming very small.

\begin{figure}
	\centering
	\includegraphics[width=6in]{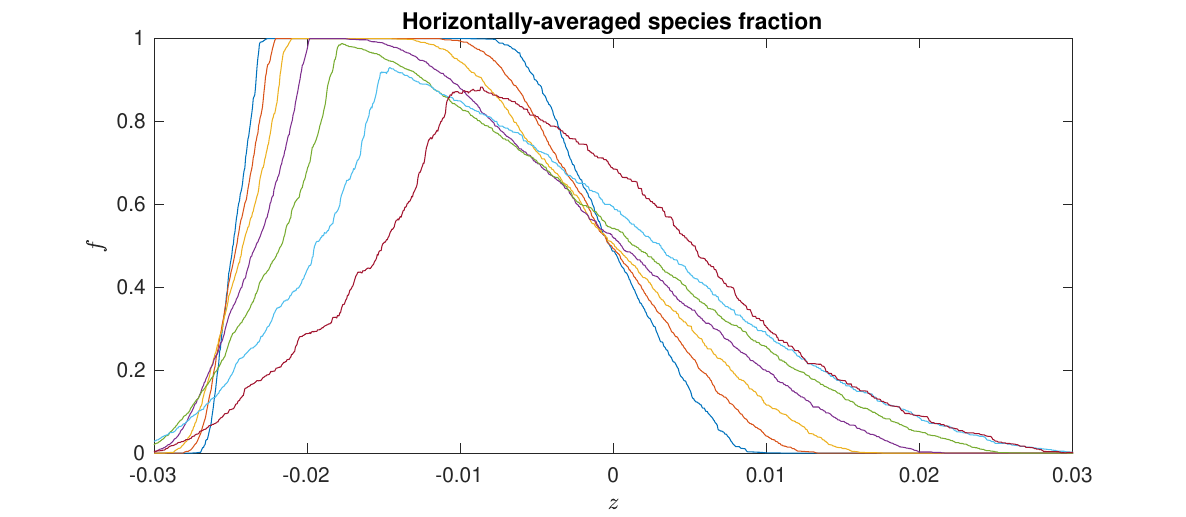}
	\includegraphics[width=6in]{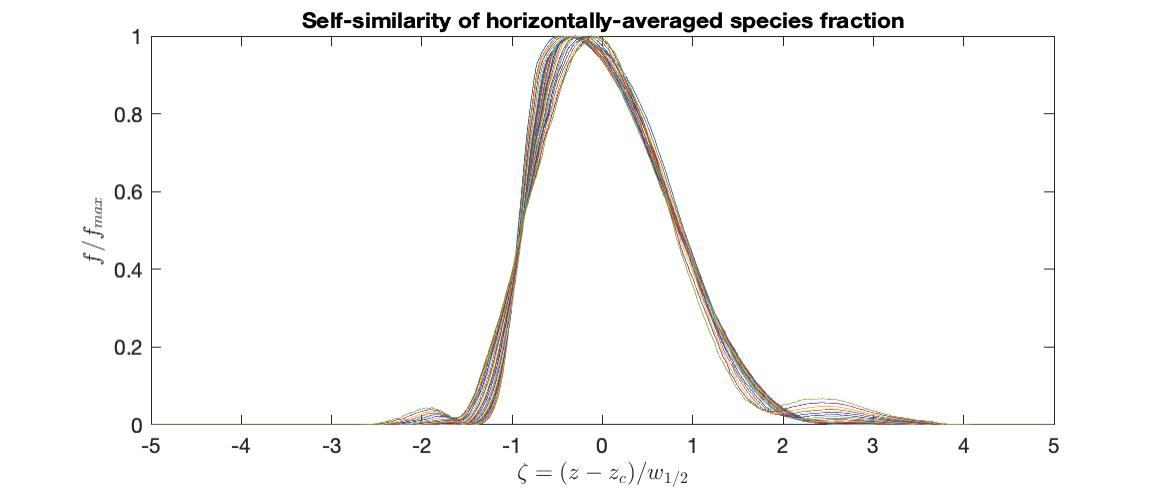}
	\caption{{\footnotesize  (\textit{Top}) Mean species fraction $f(\cdot,t)$, plotted at various times. (\textit{Bottom}) Mean species fraction, plotted in rescaled coordinates to highlight self-similarity of concentration profiles. }}
	\label{stratification}
\end{figure}

\begin{figure}
	\centering
	\includegraphics[width=5in]{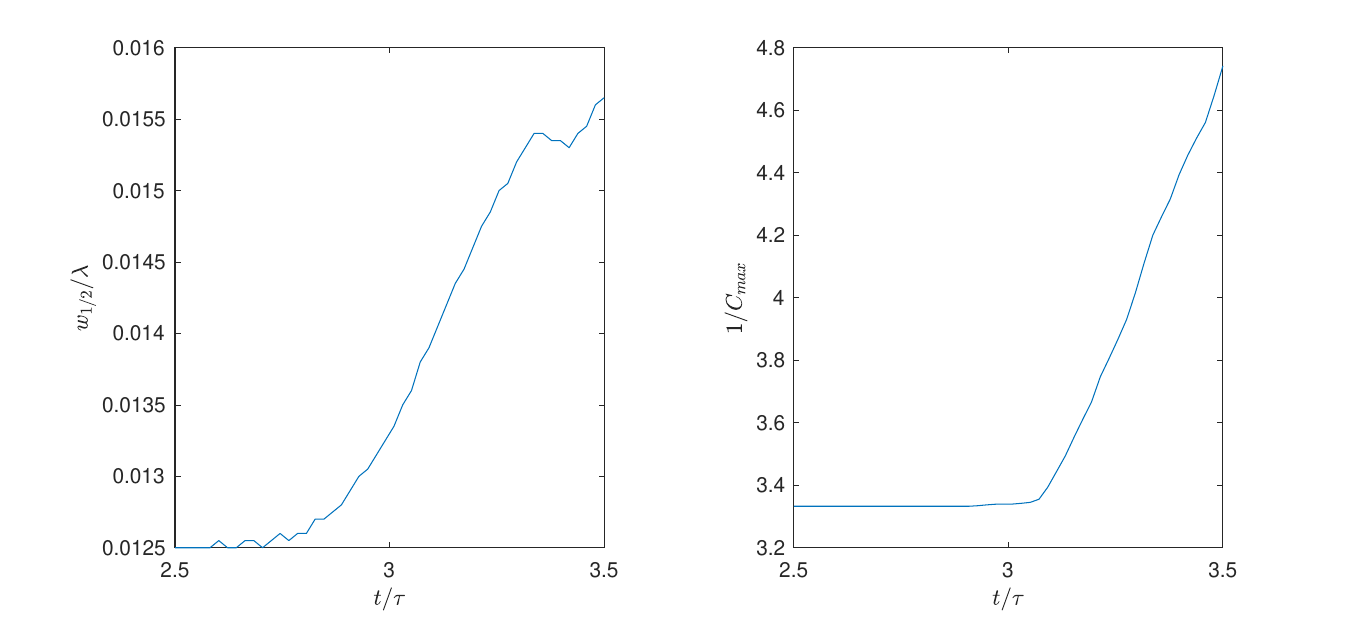}
	\caption{{\footnotesize Time evolution of the width-at-half-height $w_{\sfrac{1}{2}}(t)$ (\textit{Left}) and the reciprocal of the maximum concentration $1/f_{max}(t)$ (\textit{Right}), showing linear growth after $t/\tau=3$.}}
	\label{widthheight}
\end{figure}

We next use our multiple-interface $z$-model to study the problem of fluid entrainment.
Of interest are the experiments of \citet{DaJa2005}, who studied fluid mixing in the case where a layer of light fluid (density $\rho^-$) is initially sandwiched between two layers of heavy fluid (density $\rho^+$). 
For short times, this results in standard RTI in the top interface and stability in the bottom interface; at long times, the turbulent flow in the mixing layer becomes strong enough to disturb the lower interface and entrain some of the heavy fluid in the bottom layer.
One of the hypotheses tested by Jacobs and Dalziel is that the horizontally-averaged species fraction of light fluid, given by
\begin{equation}
	f(z,t)=\int_{\mathbb{R}^2}\frac{\rho^+-\rho(x,y,z,t)}{\rho^+-\rho^-}dxdy,
	\label{eqn::avg}
\end{equation}
exhibits an asymptotically self-similar distribution at large times, analogous to the self-similarity of the density field shown in figure \ref{rr_selfsimilar}. \footnote{Note that the species fraction differs from the density by an affine transformation. We simply choose to model the species fraction in this section to match the experimental results of \citet{DaJa2005}.}
However, unlike for the standard rocket rig, the simple Prandtl closure model will not suffice, erroneously predicting vertically symmetric growth in the concentration profile. 
Instead, one finds that 
$$\frac{f(z,t)}{f_{max}(t)} = F(\zeta),\qquad\qquad\zeta=\frac{z-z_c(t)}{w_{\sfrac{1}{2}}(t)},$$
where $z_c(t)$ is the centroid of the distribution $f(z,t)$, $w_{\sfrac{1}{2}}(t)$ is the width-at-half-height of $f(z,t)$, and $f_{max}(t)=\max f(\cdot,t)$. 
Moreover, it was shown by dimensional arguments and verified experimentally by \citet{DaJa2005} that 
$$f_{max}\sim \frac{1}{t},\qquad\qquad w_{\sfrac{1}{2}}\sim t$$
as $t\to\infty$. 
We use our multi-interface $z$-model to compute the concentration curves $f(z,t)$. 
In particular, we compute the ensemble-averaged density field from 25 $z$-model runs, and compute the horizontally-averaged species fraction using the formula \eqref{eqn::avg}.   
A selection of the resulting species fraction curves $f(z,t)$, as well as their rescaled counterparts, are shown in Figure \ref{stratification}. 
As can be seen in Figure \ref{widthheight}, solutions of the $z$-model verify the long-time growth prediction for the
functions $w_{\sfrac{1}{2}}(t)$ and $1/f_{max}(t)$ given by  Jacobs and Dalziel.  
As seen in Figure \ref{widthheight}, our computation of the functions $w_{\sfrac{1}{2}}(t)$ and $1/f_{max}(t)$ indeed demonstrate long-time linear growth. 
Thus the presence of a layer of heavy fluid below an unstably stratified density field serves to stabilize the resulting Rayleigh-Taylor instability, resulting in linear rather than quadratic growth of the mixing layer.


\section{Conclusions}

We have derived and tested our interface $z$-model for Rayleigh-Taylor instability where the material interface is represented by a parametrized surface in 3D, and the velocity is reconstructed using a boundary integral method, by assuming potential flow in the fluid bulk.
Using our models, we have studied several classical problems in Rayleigh-Taylor instability, including single-mode and random multi-mode  initial data.
We have demonstrated that ensemble averages of inviscid RTI with random multi-mode initial data reproduce the self-similar vertical density distribution predicted by a simple Prandtl closure model, and achieves the theoretically optimal exponential mixing rate for passive scalar fields. 
Additionally, the derivation of our one-interface model has been generalized to allow for multiple fluid interfaces, thus permitting the study of unstably stratified fluid flow.  
We have demonstrated self-similarity of the species fraction profile for a heavy-light-heavy initial distribution of densities, and corroborated the similarity exponents measured experimentally by \citet{DaJa2005}. 
These three-phase fluid problems provide a significant numerical challenge for conventional numerical methods, as the distance between the two interfaces becomes extremely small and the gradient of density becomes extremely large, which our model is able to simulate stably.
The idea of dimension reduction via the Birkhoff-Rott integral is now classical, but after more than six decades of simulation efforts, few competitive results exist for three-dimensional dynamics. 
We have been able to identify the appropriate modifications for the evolution of the amplitude of vorticity which allow our interface model to avoid severe numerical difficulties and retain high accuracy of the location of vorticity spikes that initial interface roll-up. 
We have demonstrated the efficacy of these models for single-mode initial data and random multi-mode initial data when compared against experimental data, which shows that two-phase potential flow is sufficiently rich to capture the complexity of 3D RTI. 
This makes our model  a stable and accurate alternative to the challenging simulations of the full two-phase irrotational and incompressible Euler equations. 
In the future, we hope to extend this model to include topological transitions in the interface, so we can test the efficacy of our model in late-stage RTI and turbulent mixing, and perform a more detailed study of optimal mixing rates in RTI-driven turbulence.


\end{document}